\documentclass[12pt,preprint]{aastex}

\shorttitle{}
\shortauthors{}

\begin{document}

\title{The Structure of the Local Interstellar Medium. II. Observations of \ion{D}{1}, \ion{C}{2}, \ion{N}{1}, \ion{O}{1}, \ion{Al}{2}, and \ion{Si}{2} toward Stars within 100 parsecs\footnote{Based on observations made with the NASA/ESA Hubble Space Telescope, obtained from the Data Archive at the Space Telescope Science Institute, which is operated by the Association of Universities for Research in Astronomy, Inc., under NASA contract NAS AR-09525.01A. These observations are associated with program \#9525.}}

\author{Seth Redfield\altaffilmark{2} and Jeffrey L. Linsky}
\affil{JILA, University of Colorado and NIST, Boulder, CO 80309-0440}
\altaffiltext{2}{Currently a Harlan J. Smith Postdoctoral Fellow at McDonald Observatory, Austin, TX 78712-0259; sredfield@astro.as.utexas.edu}

\begin{abstract}

Moderate and high-resolution measurements ($\lambda/\Delta\lambda\,\gtrsim\,$40,000) of interstellar resonance lines of \ion{D}{1}, \ion{C}{2}, \ion{N}{1}, \ion{O}{1}, \ion{Al}{2}, and \ion{Si}{2} (hereafter called light ions) are presented for all available observed targets located within 100~pc which also have high-resolution observations of interstellar \ion{Fe}{2} or \ion{Mg}{2} (heavy ions) lines.  All spectra were obtained with the Goddard High Resolution Spectrograph (GHRS) or the Space Telescope Imaging Spectrograph (STIS) instruments aboard the Hubble Space Telescope (HST).  Currently, there are 41 sightlines to targets within 100\,pc with observations that include a heavy ion at high resolution and at least one light ion at moderate or high resolution.  We present new measurements of light ions along 33 of these sightlines, and collect from the literature results for the remaining sightlines that have already been analyzed.  For all of the new observations we provide measurements of the central velocity, Doppler width parameter, and column density for each absorption component.  We greatly increase the number of sightlines with useful LISM absorption line measurements of light ions by using knowledge of the kinematic structure along a line of sight obtained from high resolution observations of intrinsically narrow absorption lines, such as \ion{Fe}{2} and \ion{Mg}{2}.  We successfully fit the absorption lines with this technique, even with moderate resolution spectra.  Because high resolution observations of heavy ions are critical for understanding the kinematic structure of local absorbers along the line of sight, we include 18 new measurements of \ion{Fe}{2} and \ion{Mg}{2} in an appendix.  We present a statistical analysis of the LISM absorption measurements, which provides an overview of some physical characteristics of warm clouds in the LISM, including temperature and turbulent velocity.  This complete collection and reduction of all available LISM absorption measurements provides an important database for studying the structure of nearby warm clouds, including ionization, abundances, and depletions.  Subsequent papers will present models for the morphology and physical properties of individual structures (clouds) in the LISM.

\end{abstract}


\section{Introduction}

Most resonance lines of important ions in the local interstellar medium (LISM) lie in the ultraviolet (UV) spectral range.  This paper continues our inventory of LISM absorption line observations observed by the two high resolution spectrographs onboard the {\it Hubble Space Telescope} ({\it HST}), the Goddard High Resolution Spectrograph (GHRS) and the Space Telescope Imaging Spectrograph (STIS).  In this paper, we extend the work of \citet{red02}, who analyzed all available LISM \ion{Fe}{2} and \ion{Mg}{2} absorption lines observed by {\it HST} and \ion{Ca}{2} lines observed from the ground, along sightlines $<$\,100\,pc, to include six more ions: \ion{D}{1}, \ion{C}{2}, \ion{N}{1}, \ion{O}{1}, \ion{Al}{2}, and \ion{Si}{2}.  These ions, which all fall in the 1200-1800\,\AA\, spectral range, extend the resonance line database to include a range of ion masses from atomic weight 2.01, \ion{D}{1}, to atomic weight 55.85, \ion{Fe}{2}.  This collection of LISM absorption line observations will form the basis for a comprehensive analysis of many of the physical properties of the LISM, which will be published in subsequent papers of this series.

The format of this work is similar to that of \citet{red02}, and we refer the reader to that paper for a thorough introduction to our procedure for the analysis of interstellar absorption lines.  The most important difference between these two papers is our inclusion of moderate resolution spectra for the analysis of low mass ions.  The velocity structure along a line of sight is best measured by resonance lines of the heaviest ions observed with high spectral resolution, as discussed by \citet{red02}.  We use the kinematic information obtained from the high resolution spectra of the high mass ions when fitting the moderate resolution spectra of the lighter ions.  Because of the cosmological and Galactic chemical evolution implications of the abundance of deuterium in the LISM \citep{lin98}, \ion{D}{1} measurements have already been published for many nearby stars, and 11 more measurements are presented in this paper.  The other light ions: \ion{C}{2}, \ion{N}{1}, \ion{O}{1}, \ion{Al}{2}, and \ion{Si}{2} have not been analyzed for the majority of sightlines with available spectra.  Although a few sightline investigations have already been presented \citep{heb99,wood97,gnac00,wrls02,vidal98}, this paper provides a valuable new database of LISM measurements of these ions which is complete to date.  By coupling the datasets presented here with those of \ion{Mg}{2} and \ion{Fe}{2} \citep{red02}, we can investigate fundamental physical properties of the LISM, including: the morphology \citep{red00}, small scale structure \citep{redfield01}, ionization structure \citep{wrls02}, and the temperature and turbulent velocity structure \citep{red03b}.

\section{Observations}

We list in Table~\ref{sw_table1} all stars located within 100 pc that have moderate ($30,000\,\leq\,\lambda/\Delta\lambda\,\leq\,45,000$) or high ($\lambda/\Delta\lambda\,\geq\,$100,000) resolution spectra of LISM absorption lines in the 1200-1800\,\AA\ spectral range.  We include in this sample only those sightlines that also have high resolution observations of \ion{Fe}{2} and/or \ion{Mg}{2} to ensure that we have an accurate measurement of the kinematic structure along the line of sight \citep{red02}.  \ion{Ca}{2} is not used alone, as a kinematic constraint on other UV lines, because low column density clouds not detected in \ion{Ca}{2} are observed in stronger UV resonance lines (e.g. the $\alpha$~Cen line of sight).  Resonance lines that show LISM absorption in the 1200-1800\,\AA\ spectral range include: the \ion{D}{1} fine structure doublet at 1215.3376\,\AA\ and 1215.3430\,\AA, \ion{C}{2} at 1334.532\,\AA, the \ion{N}{1} triplet at 1199.5496\,\AA, 1200.2233\,\AA, and 1200.7098\,\AA, \ion{O}{1} at 1302.1685\,\AA, \ion{Al}{2} at 1670.7874\,\AA, and five \ion{Si}{2} resonance lines at 1190.4158\,\AA, 1193.2897\,\AA, 1260.4221\,\AA, 1304.3702\,\AA, and 1526.7066\,\AA.  Because some sightlines with light ion absorption spectra included recent high resolution observations of \ion{Mg}{2} and \ion{Fe}{2}, we have included the analysis of new heavy ion absorption lines in Appendix~\ref{app}.  These measurements complement the \ion{Mg}{2} and \ion{Fe}{2} survey published by \citet{red02}.

For all of the stars, the stellar spectral type, visible magnitude, stellar radial velocity, and Galactic coordinates are taken from the SIMBAD database unless otherwise noted.  The {\it Hipparcos} distances \citep{perry97} are listed without errors, but the errors in trigonometric parallax measurements for stars within 100\,pc are small enough to have no influence on the interpretation of our results.  In general, at 1~pc the $1\sigma$ error in the distance is $\sim~0.1$~pc, at 30~pc the $1\sigma$ error is $\sim~1.0$~pc, and at 100~pc the $1\sigma$ error is $\sim~5.0$~pc.  Also listed are the predicted absorption velocities for the Local Interstellar Cloud (LIC) and Galactic (G) clouds computed using the vectors proposed by \citet{lall92} and \citet{lall95}.

All observations were taken with the high resolution spectrographs onboard the {\it HST}:  the GHRS and the STIS instruments.  The GHRS instrument is described by \citet{brandt94} and \citet{heap95}.  \citet{soder94} describe improvements to the GHRS instrument for observations taken after the installation of the Corrective Optics Space Telescope Axial Replacement (COSTAR) in December~1993.  The STIS instrument is described by \citet{kimble98} and by \citet{woodgate98}.

Many {\it HST} observations of the stars listed in Table~\ref{sw_table1} were taken for the express purpose of studying the structure of the LISM along these sight lines.  However, more than a third of the observations along lines of sight listed in Table~\ref{sw_table1} were taken for other purposes.   We list in Table~\ref{sw_table2} the {\it HST} observations of those stars with LISM absorption data that have not been published.  All of these data were taken from the {\it HST} Data Archive and are publicly available.  The data reduction procedure for the observations listed in Table~\ref{sw_table2} is described in Section~\ref{sw_datareduc}, and the spectral analysis is discussed in Section~\ref{sw_specanal}.

\subsection{Data Reduction \label{sw_datareduc}}

We reduced the GHRS data acquired from the {\it HST} Data Archive with the CALHRS software package using the Image Reduction and Analysis Facility (IRAF) and the Space Telescope Science Data Analysis System (STSDAS) software.  The most recent reference files were used in the reduction.  In many cases, the current reference files produced an improvement in data quality, compared to data reduced with reference files contemporaneous with the observations.  Many of the echelle observations were obtained in the FP-SPLIT mode to reduce fixed-pattern noise.  The individual readouts of the FP-SPLIT spectra were combined using a cross-correlation procedure called HRS\_MERGE \citep{robinson92}.  The reduction included assignment of wavelengths using calibration spectra obtained during the course of the observations.  The calibration spectra include either a WAVECAL, a direct Pt-Ne lamp spectrum from which a dispersion relation can be obtained, or a SPYBAL (Spectrum Y-Balance), from which only a zero-point offset can be obtained.  Any significant errors involved in the wavelength calibration are included in our central velocity determinations.

We reduced the STIS data acquired from the {\it HST} Data Archive using the STIS team's CALSTIS software package written in IDL \citep{lindler99}.  The reduction included assignment of wavelengths using calibration spectra obtained during the course of the observations.  The ECHELLE\_SCAT routine in the CALSTIS software package was used to remove scattered light.  

\subsection{Spectral Analysis \label{sw_specanal}}

Figures~\ref{sw_fig1}-\ref{sw_fig6} show our fits to the LISM absorption lines that were observed in the 1200-1800\,\AA\ spectral range and have not yet been published (see Table~\ref{sw_table2} for a list of the observational parameters).  All of the spectra are shown with a heliocentric velocity scale, together with our estimate of the stellar continuum ({\it thin solid lines}), the best fits to the absorption by each interstellar component ({\it dashed lines}), and the total interstellar absorption convolved with the instrumental profile ({\it thick solid lines}).  Table~\ref{sw_table2_5} presents an inventory of detected LISM absorption toward all 41 stars within 100\,pc that have both a high resolution heavy ion detection and at least one moderate to high resolution detection of a light ion.  We differentiate the new measurements presented in this work, with those already analyzed in the literature.  Of the 41 sightlines listed in Table~\ref{sw_table2_5}, 33 have new LISM absorption line observations.

The LISM absorption lines are fit using standard techniques \citep{lin96,pisk97,dring97,red02}.  This involves estimating the stellar continuum, determining the fewest number of absorption components required to obtain a satisfactory fit, convolving the absorption feature with an instrumental line spread function, and fitting all lines of the same ion simultaneously.  Determining the stellar continuum is usually straightforward when dealing with high resolution data, and systematic errors in this procedure are further reduced when several resonance lines of the same ion are fit simultaneously.  Because the central wavelength of the interstellar absorption is different for each line of sight, the placement of the unobserved stellar flux for some cases can be both difficult and the dominant source of error.  In most cases, the ``continuum'' against which the absorption is measured can be determined by fitting a simple polynomial to spectral regions just blueward and redward of the interstellar absorption.  Of the 206 interstellar absorption spectra shown in Figures~\ref{sw_fig1}-\ref{sw_fig6} (as well as those shown in Appendix~\ref{app}), 68\% of the continua were determined in this straightforward way.  

Occasionally, a simple polynomial is not sufficient to accurately estimate the missing stellar continuum.  We employ three additional techniques to accurately estimate the continuum: mirroring, similarity, and simultaneous fitting.  The mirroring technique for estimating the unobserved stellar flux involves reflecting the emission line about the stellar photospheric radial velocity.  If the interstellar absorption is far from line center, typically because of the high radial velocity of the star, this technique can be very useful in estimating the stellar flux.  Since these circumstances are not very common, we used this method for estimating the continuum for only 5\% of the sample.  Fitting the \ion{C}{2} line of 36~Oph~A and 70~Oph (see Figure~\ref{sw_fig1}) are examples of this technique.  

A stellar emission line with LISM absorption is often a member of a multiplet.  Because the other members of the multiplet are not resonance transitions, no LISM absorption is seen against these emission lines.  The shape of this line can be used to help reconstruct the continuum of the absorbed resonance line.  This is particularly useful with lines that may or may not exhibit self-reversals at line center or multiple Gaussian profiles, such as \ion{Mg}{2}, \ion{C}{2}, and \ion{O}{1}, as well as lines that are severely blended with additional lines or a rising stellar continuum, such as \ion{Fe}{2}.  Although the line width and integrated flux vary among multiplet members, the shapes of the line profiles are often very similar.  Examples of this technique are the \ion{C}{2} lines of $\epsilon$~Eri (see Figure~\ref{sw_fig1}), and the \ion{O}{1} lines of $\beta$~Gem, $\beta$~Cet, and $\iota$~Cap (see Figures~\ref{sw_fig2}, \ref{sw_fig3}, and \ref{sw_fig5}, respectively).  This method is used for estimating 13\% of the continua in this sample.  

One of the strengths of this sample, is that there are often several absorption lines for the same ion, particularly \ion{N}{1}, \ion{Si}{2}, \ion{Mg}{2}, and \ion{Fe}{2}.  Because the oscillator strengths are known, the measured absorption in one line, predicts the absorption for all of the other resonance lines of the same ion.  In this way, we can estimate continua for low signal-to-noise (S/N) absorption features in moderate resolution data for an ion, as long as we have at least one other reasonably simple absorption feature of the same ion.  By simultaneously fitting all of the absorption features of the same ion, we can make a reasonable estimate of the continuum that is consistent with each absorption line.  Although this technique can cause some curious continua estimates, it is the best method for reducing the systematic errors involved in continua placement.  In only a few cases, approximately 14\%, does this method produce results that deviate from a simple polynomial continuum estimation.  Some examples include the \ion{Si}{2} lines of GD~153 (see Figure~\ref{sw_fig5}) and the \ion{Fe}{2} lines of HD~184499.  

Once the continuum has been determined, we apply the minimum number of absorption components for a satisfactory fit to the spectra.  Adding more absorption components invariably improves the goodness-of-fit metric ($\chi^2$).  We use the F-test to determine whether the improvement in the fit is statistically significant \citep{bev92}, and allows us to determine the minimum number of necessary absorption components in an objective way.  

The instrumental line spread functions assumed in our fits for GHRS spectra are taken from \citet{gill94} and for STIS spectra from \citet{sahu99}.  The rest wavelengths and oscillator strengths of the lines used in our fits are taken from \citet{mort91} with updates from D.C. Morton (2003, private communication).  We fit all lines of each ion simultaneously in order to obtain the most robust parameters for the interstellar absorption.  The difference in oscillator strengths between the various lines can be very useful in constraining the interstellar parameters, the stellar continuum level, and the number of absorption components.  By providing an independent measure of the same absorbing column using absorption lines with different optical depths and at different locations relative to the stellar continuum, our analysis of several lines of the same ion reduces the effect of systematic errors, in particular from saturation and line blends, that result from the measurement of only one line.

Occasionally, absorption features due to terrestrial material are detected in spectra of \ion{O}{1} and \ion{N}{1} (see HZ\,43, GD\,153, and DK\,UMa, in Figures~\ref{sw_fig4}-\ref{sw_fig6}).  The absorption is easily identified because it is centered at the mean velocity of the Earth during the time of observation, and because it is often also seen in the other members of the multiplet, where no LISM absorption is detected.  The airglow absorption features are easily modeled and do not adversely affect the fit to the LISM absorption lines.  

For each absorption component there are three fit parameters: the central velocity ($v$ [km~s$^{-1}$]), the Doppler width ($b$ [km~s$^{-1}$]), and the column density ($ N_{\rm ion}$ [cm$^{-2}$]).  The central velocity corresponds to the mean projected velocity of the absorbing material along the line of sight to the star.  If, as a first approximation, we assume that the warm partially ionized material in the solar neighborhood exists in small, homogeneous cloudlets each moving with a single bulk velocity, then each absorption component will correspond to a single cloudlet, and its velocity is the projection of its three-dimensional velocity vector along the line of sight.  The Doppler width parameter is related to the temperature ($T$ [K]) and nonthermal velocity ($\xi$ [km~s$^{-1}$]) of the interstellar material by the following equation:
\begin{equation}
b^2 = \frac{2kT}{m}+\xi^2 = 0.016629\left(\frac{T}{A}\right)+\xi^2 , 
\label{sw_eq1}
\end{equation}
where $k$ is Boltzmann's constant, $m$ is the mass of the ion observed, and $A$ is the atomic weight of the element in atomic mass units (where $A$ can range from $A_{\rm D}=2.01410$ to $A_{\rm Fe}=55.847$).  The Doppler width parameter, or line width, of an ion with a large atomic weight is more sensitive to turbulent broadening than thermal broadening.  The column density measures the amount of material along the line of sight to the star.  If we assume homogeneous cloudlets with constant density, then the column density will be directly proportional to the cloud thickness.  The compilation of all measurements for stars within 100~pc provides an important database for the analysis of the structure of the LISM.  The velocity structure is discussed in Section~\ref{sw_veldist}, the Doppler width parameter structure is discussed in Section~\ref{sw_doppdist}, and the column density structure is discussed in Section~\ref{sw_coldist}.

In order to extract as much information as possible from the UV observations of LISM absorption lines, we apply knowledge of the velocity structure along the line of sight derived from high resolution spectra of intrinsically narrow absorption lines of \ion{Fe}{2} and \ion{Mg}{2} to the fitting of the moderate resolution spectra of the lower mass, and thus, intrinsically broader absorption lines.  For example, an \ion{Fe}{2} absorption line formed by interstellar material at the canonical LISM temperature, $\sim\,7000$\,K, and with no turbulent broadening will have a Doppler line width of $b\,=\,1.44$\,km\,s$^{-1}$, whereas the \ion{D}{1} absorption line for the same conditions will have $b\,=\,7.60$\,km\,s$^{-1}$.  To fully resolve the interstellar \ion{Fe}{2} absorption feature requires a spectral resolution of $R\,=\,\lambda/\Delta\lambda\,>\,$200,000, whereas the \ion{D}{1} feature requires a resolution of $R\,>\,$40,000.  If the heavy ions accurately trace the same collection of gas as the light ions, then the required resolving power can decrease, as the atomic weight of the ion decreases, provided that the kinematic structure along the line of sight has already been determined from the analysis of intrinsically narrow lines observed at high resolution.  

In this paper, we take advantage of such high resolution observations of \ion{Fe}{2} and \ion{Mg}{2} lines presented by \citet{red02} to fit moderate resolution data of the same sightlines.  In particular, we fit the same number of velocity components to the light ion spectra that were necessary in the high resolution spectra.  In practice, the vast majority of light ion absorption spectra have enough spectral resolution and signal-to-noise to independently constrain the number of components and their central velocities.  However, in cases where the absorption components are severely blended, or the S/N is particularly poor, we use information from the high resolution fits to \ion{Fe}{2} and \ion{Mg}{2}.  In only two sightlines, 70~Oph (\ion{C}{2}, \ion{O}{1}, and \ion{Si}{2}) and HD~28568 (\ion{C}{2}), the absorption features are so blended that we use the velocity information directly from the \ion{Mg}{2} data to fit these spectra.  In a handful of additional cases, the absolute velocity structure is not constrained, only the differences in central velocities are taken from the high resolution fits to constrain the light ion spectra.

Tables~\ref{sw_table3} through \ref{sw_table8} list the interstellar absorption parameters and $1\sigma$ errors for all targets in Table~\ref{sw_table1} that show interstellar absorption in \ion{D}{1}, \ion{C}{2}, \ion{N}{1}, \ion{O}{1}, \ion{Al}{2}, and \ion{Si}{2}, respectively.  The data in Tables~\ref{sw_table3}-\ref{sw_table8} include our fits to the lines of sight shown in Figures~\ref{sw_fig1}-\ref{sw_fig6}.  All errors quoted in the text, tables, and figures are $1\sigma$ error bars.  In Table~\ref{sw_table3} we list 1$\sigma$ errors in $\log N_{\rm DI}$ that for the best cases are as small as 0.01. For those lines of sight with only one velocity component and high S/N spectra, we can reliably interpolate the \ion{H}{1} line profile in the neighborhood of the \ion{D}{1} line to achieve this precision.  Observations with moderate S/N, and/or several velocity components detected along the line of sight, typically have uncertainties in $\log N_{\rm DI}$ on the order of 0.1.

\section{Discussion}

\begin{figure}
\epsscale{.9}
\plotone{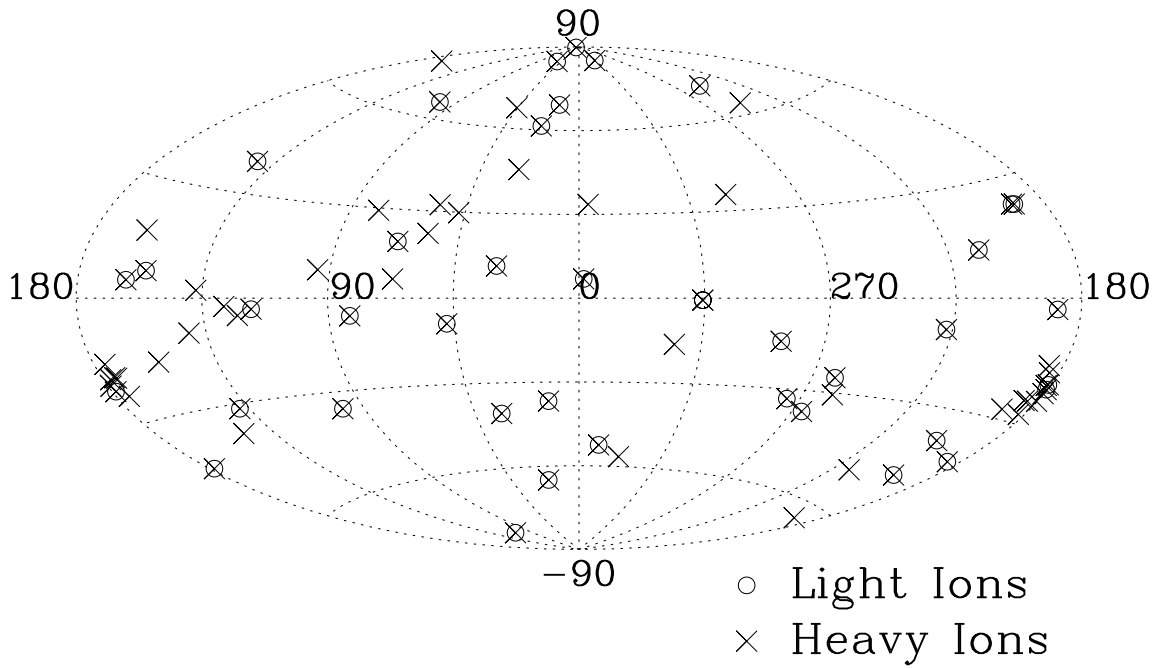}
\caption{The location of all stars listed in Table~\ref{sw_table1} are shown in Galactic coordinates.  All {\it HST} targets with high resolution spectra of \ion{Mg}{2} and/or \ion{Fe}{2} are indicated by the ``$\times$'' symbol.  There are 80 sightlines within 100\,pc, which currently have these heavy ion detections (see \citet{red02} and Appendix~\ref{app} of this work).  Those stars that in addition, have moderate or high resolution spectra of any light ions, such as \ion{D}{1}, \ion{C}{2}, \ion{N}{1}, and \ion{O}{1}, are plotted as open circles.  Of the 41 sightlines that have both high resolution observations of a heavy ion and at last one observation of a light ion at moderate or high resolution, new data are presented in this work for 33 lines of sight.  
\label{sw_fig9}}
\end{figure}

The following discussion will be primarily a statistical analysis of the LISM absorption fit parameters discussed above.  In future papers we will present more detailed analyses of the database, including the morphology, kinematics, and physical properties of identifiable structures (i.e. clouds) in the LISM.  The location of all stars listed in Table~\ref{sw_table1} are shown in Galactic coordinates in Figure~\ref{sw_fig9}.  Although spatial coverage overall is fairly good, there are areas of the sky that are poorly sampled, while others are densely sampled.

\subsection{Velocity Distribution \label{sw_veldist}}

Because the focus of this paper is the fitting of relatively broad absorption lines of low mass ions with moderate resolution data, the measurement of velocity centroids is unlikely to improve the accuracy obtained from fits to the high resolution spectra of narrow absorption lines.  However, our analysis of the low mass ion profiles provides an opportunity to measure the consistency of the observed velocity centroids for a particular absorber.  By comparing the measured absorption velocities of many resonance lines along the same sightline, we can test: (1) the implicit assumption that all of the observed ions sample the same collection of gas, and will therefore have the same observed projected velocity, and (2) the magnitude of systematic errors involved in comparing observations obtained with different spectral resolutions and observed with different instruments (e.g. GHRS or STIS).  Ionization models of the LISM by \citet{slavin02} indicate that all of the measured ions discussed in this paper are expected to be the dominate ion of the element, and therefore we expect them all to sample the same collection of gas.  

\begin{figure}
\epsscale{.9}
\plotone{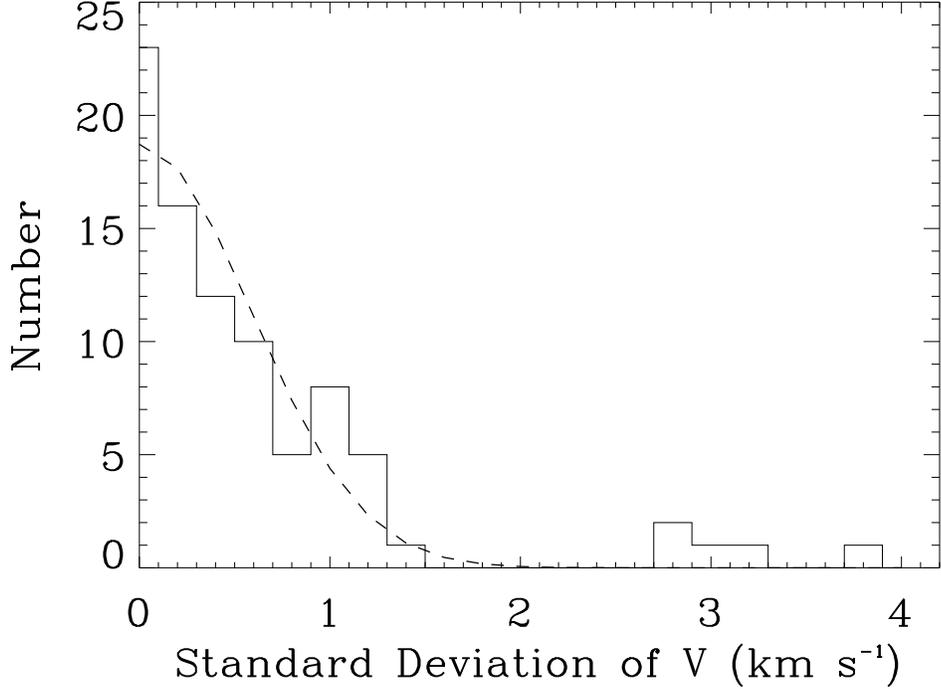}
\caption{The distribution of standard deviations (the square root of the weighted average variance) about the weighted mean velocity centroid measurements in resonance lines of different ions for individual LISM components along the same line of sight.  There are 87 individual components along 45 sightlines, where two or more resonance lines trace absorption from the same collection of local gas.  The sightlines include those presented in this paper with both heavy and light ion observations, plus a few lines of sight with both \ion{Fe}{2} and \ion{Mg}{2}, but no light ion observations, as discussed by \citet{red02} and in Appendix~\ref{app}.  Each absorption line provides an independent measurement of the projected velocity of the absorbing gas.  However, due to random and systematic errors, those measurements are distributed about the true value.  This figure shows the standard deviation about the weighted mean for all of the velocity centroid measurements for each component along each line of sight.  The dashed line indicates a Gaussian distribution with a mean of 0\,km\,s$^{-1}$, and a standard deviation of 0.59\,km\,s$^{-1}$.  The large majority (78\%) of velocity components agree to within 1.0~km~s$^{-1}$. The outliers are most likely caused by unresolved blends, or large systematic errors in the instrumental wavelength calibration.  The bin size used is 0.2~km~s$^{-1}$.\label{sw_fig10}}
\end{figure}

Each individual sightline now includes multiple measurements with different ions of the projected velocity of each interstellar absorber (cloud) along the line of sight.  Figure~\ref{sw_fig10} shows the distribution of the standard deviation about the weighted mean of these velocity measurements for each particular absorber.
If all ions sample the same collection of gas and all of our instruments were infinitely precise and perfectly calibrated, then the standard deviation should be $\sim\,0$.  Although there is a distribution of standard deviations, it is strongly peaked at 0.  Therefore, all of the measured ions for each cloud appear to sample the same collection of gas.  The dashed line indicates a Gaussian fit to the distribution, which well characterizes the data.  The Gaussian has a standard deviation of $\sigma_{\rm G}\,=\,0.59\,$km\,s$^{-1}$.  We believe that this is a good estimate of the typical systematic error involved in measuring LISM absorption lines in moderate and high resolution spectra observed with the GHRS and STIS instruments.  Because we are comparing a large number of individual observations, each with its own systematic errors particular to each absorption feature (i.e. continuum placement, degree of blending with other lines, etc...), the distribution of systematic errors approaches a random distribution.  Therefore, the range of standard deviations for all of the central velocity measurements of a particular absorber (cloud) should and does approximate a Gaussian distribution with a rather small half width.

\subsection{Doppler Width Parameter Distribution\label{sw_doppdist}}

\begin{figure}
\epsscale{.69}
\plotone{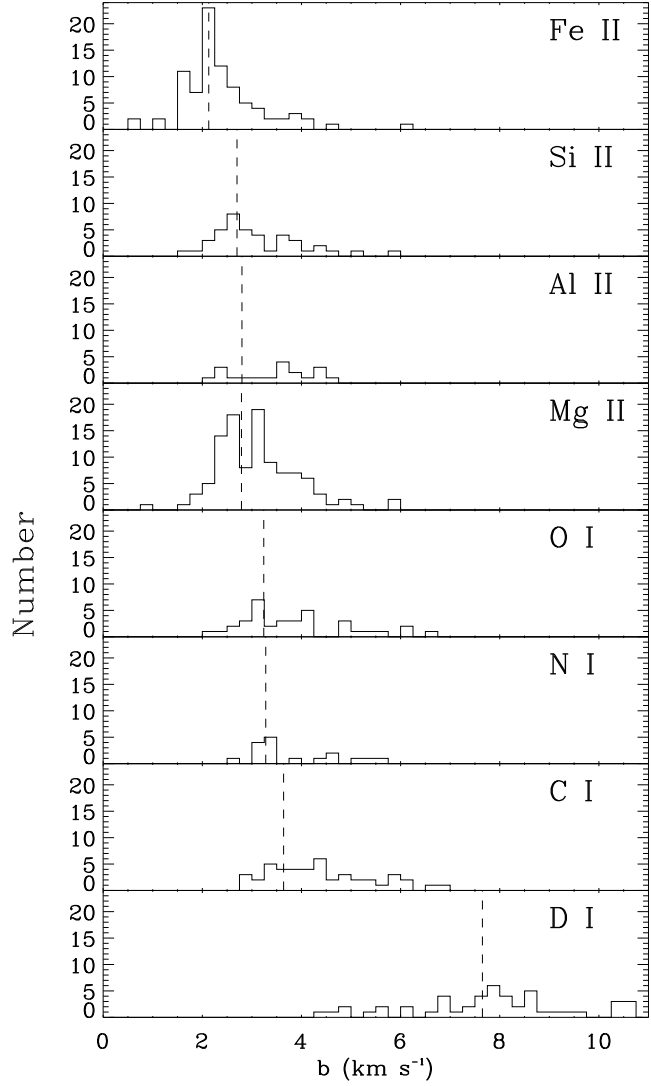}
\caption{The Doppler width parameter distribution for eight ions, from the heaviest ion, \ion{Fe}{2}, at the top, to the lightest ion, \ion{D}{1}, at the bottom.  The bin size used is 0.25~km~s$^{-1}$.  The dashed lines indicate the weighted mean Doppler width.  The weighted mean, standard deviation of the weighted mean, and the total number of observations (N) of each ion are presented in Table~\ref{sw_table11}.  A clear shift to larger line widths is seen toward lighter ions, indicating that for the heaviest ions turbulent broadening dominates and the Doppler width parameter is independent of atomic mass, while for the lightest ion thermal broadening dominates.  \label{sw_fig11}}
\end{figure}

The observed value of the Doppler width parameter ($b$) is a measure of the temperature ($T$) and turbulent velocity ($\xi$).  The relationship is provided by Equation~\ref{sw_eq1}.  With increasing atomic mass, the Doppler width parameter is influenced less by thermal broadening and more by turbulence or unresolved clouds along the line of sight.  In Figure~\ref{sw_fig11}, the Doppler width parameter distribution is plotted for all ions, from the heaviest ion, \ion{Fe}{2}, at the top, to the lightest ion, \ion{D}{1}, at the bottom.  The increase in line width for ions with decreasing atomic weight demonstrates that thermal broadening becomes increasingly important relative to turbulent broadening for the lightest ions.   The Doppler width parameter distributions of the heaviest ions look very similar, probably because turbulent broadening completely dominates over the thermal broadening, thereby making the Doppler width parameter independent of atomic mass.  The weighted mean and standard deviation of the Doppler width parameter for all of the ions is given in Table~\ref{sw_table11}.  The high Doppler width parameter tails for all of the distributions likely results from unresolved blends \citep{welty96}.  
We attempt to minimize the occurrence of unresolved blends by, (1) observing many absorbing ions along the same line of sight, (2) observing optically thin absorption lines where individual cloud absorbers are more easily identified, (3) using high resolution observations, and (4) observing the shortest ($\leq\,$100\,pc) lines of sight to keep the absorption profile relatively simple.  However, despite our efforts, we may not be resolving all the absorbers.
Observations with higher spectral resolution would help to disentangle the component structure and to fully resolve the smallest line widths, allowing us to investigate more accurately the thermal and nonthermal characteristics of warm clouds in the LISM.  

\begin{figure}
\epsscale{.9}
\plotone{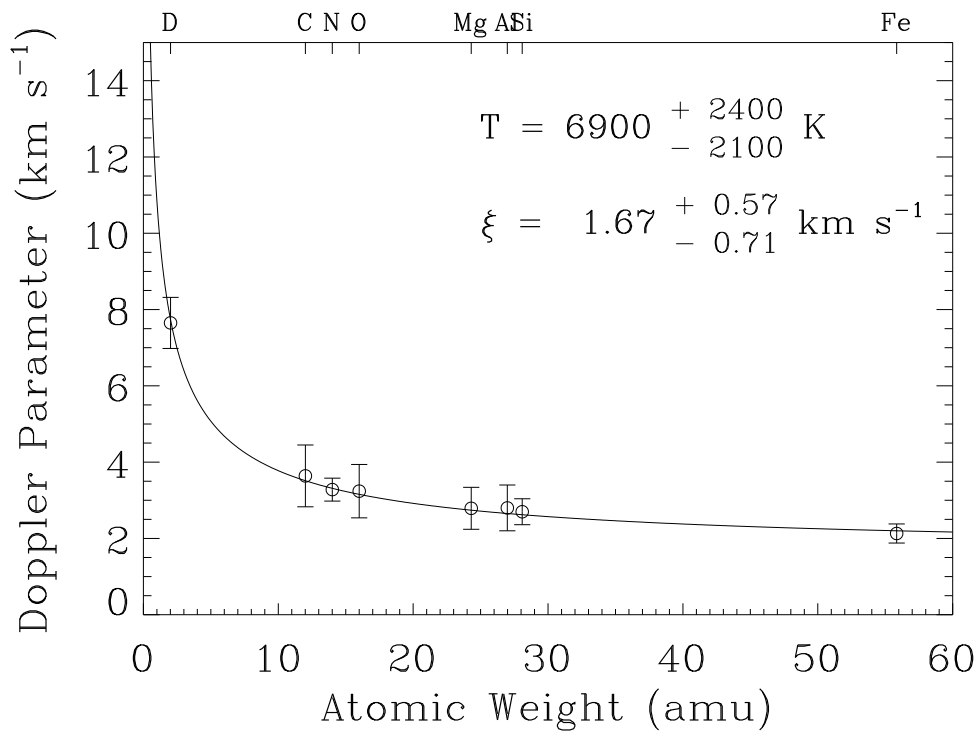}
\caption{The weighted mean Doppler width parameter for all LISM absorbers, is plotted as a function of the atomic weight of the observed ion.  The error bars indicate the weighted standard deviation of the sample.  The ions are identified along the top axis.  The best fit temperature and turbulent velocity for the LISM in aggregate is given in the plot, and the best fit curve is displayed by the solid line, as described by Equation~\ref{sw_eq1}.  Variations in temperature and turbulent velocity do exist among the various warm clouds in the LISM, as discussed in detail in \citet{red03b}. \label{sw_fig11_5} }
\end{figure}

Comparing the line widths of various atomic species with a range of atomic masses, is a powerful technique for measuring the temperature and turbulent velocity of clouds in the LISM.  In Figure~\ref{sw_fig11_5}, we plot the mean Doppler width parameters, inversely weighted by the variance of each measurement, of the entire dataset, as a function of atomic weight.  We fit the weighted mean values to Equation~\ref{sw_eq1} to calculate a typical temperature and turbulent velocity for the LISM in the aggregate.  The variations in temperature and turbulent velocity among the various warm clouds in the LISM will be discussed in detail in \citet{red03b}.  As a whole, the LISM can be roughly characterized by a temperature of $6900^{+2400}_{-2100}\,$K and a turbulent velocity of $1.67^{+0.57}_{-0.71}$\,km\,s$^{-1}$, where the error bars likely reflect the intrinsic range in these parameters rather than measurement uncertainties \citep{red03b}.

\subsection{Column Density Distribution\label{sw_coldist}}

The observed column densities give the number of ions along the line of sight to the background source.  If we assume a homogeneous medium with constant gas density, the column density is directly proportional to the distance that the line of sight traverses through the cloud.  If the gas density is known, the column density of an individual absorption component is a direct measurement of the cloud thickness, which is vital to understanding the morphology of the LISM.  The measured abundances also provide information on the degree of ionization and depletion of ions in the LISM.  These characteristics can vary from cloud to cloud and even within a single cloud structure.

\begin{figure}
\epsscale{.69}
\plotone{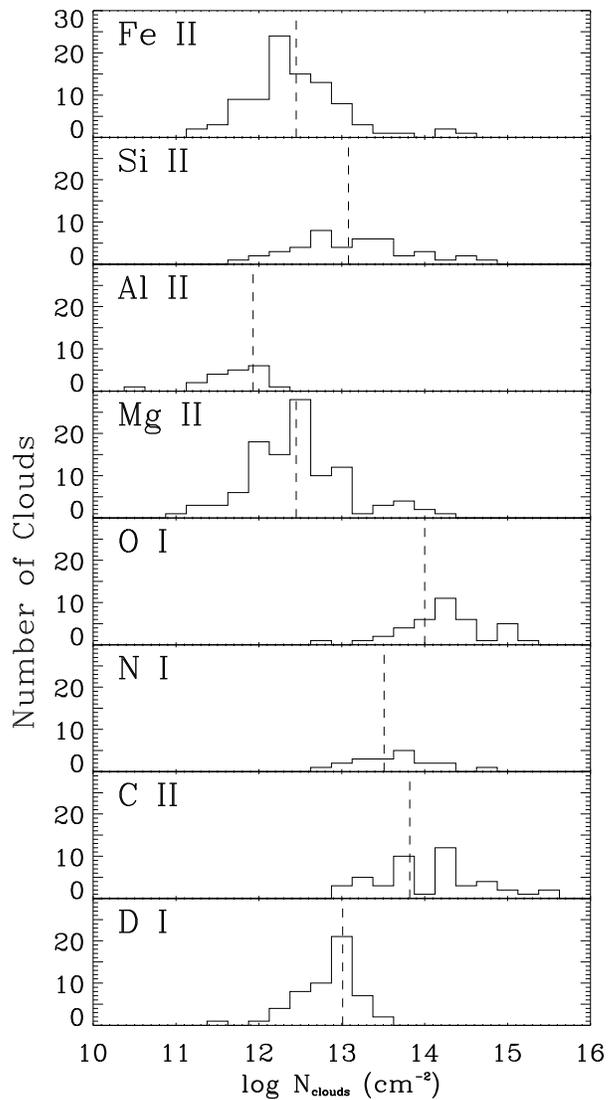}
\caption{The column density distribution for individual clouds is given for each ion.  The dashed lines indicate the weighted mean column densities.  The weighted mean, standard deviation of the weighted mean, and the total number of observations (N) of each ion are presented in Table~\ref{sw_table11}.  The variations among the various ions results from different cosmic abundances, depletions, and ionization structure.  \label{sw_fig18}}
\end{figure}

The cloud column density distribution function for each ion is given in Figure~\ref{sw_fig18}.  The distribution functions vary due to differences in cosmic abundances, depletion patterns, and ionization structure.  The weighted mean and standard deviation of the individual cloud column density for all of the ions are given in Table~\ref{sw_table11}.  The relatively small dispersion in the column densities, particularly in \ion{D}{1},  suggests that nearby clouds in the LISM may have similar sizes.  If we assume a constant D/H ratio of $1.5\,\times\,10^{-5}$ in the LISM \citep{lin98,hwm02}, and a constant H\,I number density of 0.1\,cm$^{-3}$ \citep{red00}, then we can use the weighted mean and standard deviation of the \ion{D}{1} column density distribution to estimate a typical length scale for clouds in the LISM.  This calculation results in a mean length of $2.2^{+1.7}_{-1.0}$\,pc, and a range from 0.1\,pc to 11\,pc.  

In Figures~\ref{sw_fig12}-\ref{sw_fig17}, the total column density for each ion is shown in Galactic coordinates.  The size of a symbol is inversely proportional to the target's distance, and the shading of a symbol indicates the total column density, as shown by the scale at the bottom of each plot.  The total column densities are clearly spatially correlated.  As was found by \citet{genova90}, the highest column densities tend to be for lines of sight south of the Galactic Center, and the lowest column densities tend to be in the northern anti-Galactic Center direction.  However, exceptions to this generalization exist.  Good spatial coverage of the sky and of objects at various distances is crucial to understanding the morphology of the LISM.  

\section{Conclusions}

We present a compilation of LISM absorption line measurements of ions in the spectral range from 1200-1800\,\AA, which includes lines of \ion{D}{1}, \ion{C}{2}, \ion{N}{1}, \ion{O}{1}, \ion{Al}{2}, and \ion{Si}{2}.  This work is a companion paper to the research presented by \citet{red02} on the high resolution LISM absorption lines of the heavy ions \ion{Fe}{2}, \ion{Ca}{2} and \ion{Mg}{2}.  These two papers provide a complete inventory of available LISM measurements in the UV that can be used to determine the physical properties of the LISM and to test models of the structure of the gas in our local environment.  The results of this work can be summarized as follows:
\begin{enumerate}
\item We take advantage of the known kinematic structure along a line of sight, obtained from high resolution observations of intrinsically narrow (high resolution) absorption lines \citep{red02}, to accurately measure the LISM absorption features of intrinsically broad (low mass ion) lines.  We successfully fit the absorption lines with this technique, even with moderate resolution spectra.  This greatly enlarges the number of sightlines with useful LISM absorption line measurements.
\item The central velocity of a particular absorbing cloud is now measured using many resonance lines of different ions.  The central velocity measurement for a given velocity component made with two different instruments (GHRS and STIS) and with a range of spectral resolutions have a standard deviation of only 0.59\,km\,s$^{-1}$.  
\item The distribution of Doppler widths shows a systematic increase as the atomic weight of the absorbing ion decreases.  This demonstrates that thermal broadening becomes more important compared to turbulent broadening with decreasing ion mass.  For the LISM in aggregate, we measure a temperature of $6900^{+2400}_{-2100}\,$K and a turbulent velocity of $1.67^{+0.57}_{-0.71}$\,km\,s$^{-1}$.  Variations in temperature and turbulent velocity are significant in the LISM as is discussed in detail in \citet{red03b}.
\end{enumerate}

\acknowledgments
Support for program \#9525 is provided to the University of Colorado at Boulder by NASA through a grant from the Space Telescope Science Institute, which is operated by the Association of Universities for Research in Astronomy, Inc., under NASA contract NAS AR-09525.01A.  This work is also supported by a NASA GSRP student fellowship grant NGT 5-50242.  We would also like to thank Tom Ayres and Mike Shull for their helpful comments and suggestions.

\appendix

\section{High Resolution Observations of \ion{Fe}{2} and \ion{Mg}{2} \label{app}}

A survey of all available high resolution observations of \ion{Fe}{2} and \ion{Mg}{2} absorption along sightlines within 100\,pc was presented by \citet{red02}.  We have expanded the survey in this work to include other ions, such as, \ion{D}{1}, \ion{C}{2}, \ion{N}{1}, \ion{O}{1}, \ion{Al}{2}, and \ion{Si}{2}.  For three lines of sight with new light ion measurements presented here, we also have new heavy ion observations that we analyzed before fitting the moderate to high resolution light ion spectra.  We include in this appendix the new high resolution \ion{Fe}{2} line (2600.1729~\AA, 2586.6500~\AA, 2382.7652~\AA, 2374.4612~\AA, 2344.2139~\AA) and \ion{Mg}{2}~h and k line (2803.5305~\AA\ and 2796.5318~\AA) observations toward stars within 100\,pc, since the publication of \citet{red02}.  These sightlines are listed in Table~\ref{sw_table_a1}, while the particular observational details are given in Table~\ref{sw_table_a2}.

Figures~\ref{sw_fig7}-\ref{sw_fig8b} show our best fits to the new \ion{Mg}{2} and \ion{Fe}{2} observations.  Tables~\ref{sw_table9} and \ref{sw_table10} give the absorption line fit parameters for \ion{Mg}{2} and \ion{Fe}{2}, respectively, for all targets shown in Figures~\ref{sw_fig7}-\ref{sw_fig8b}.  In addition, some errors in the tables of \citet{red02} are corrected here.  Tables~\ref{sw_table9} and \ref{sw_table10} are meant to be continuations of Tables~3 and 4 in \citet{red02}.

\begin{figure}
\epsscale{.9}
\plotone{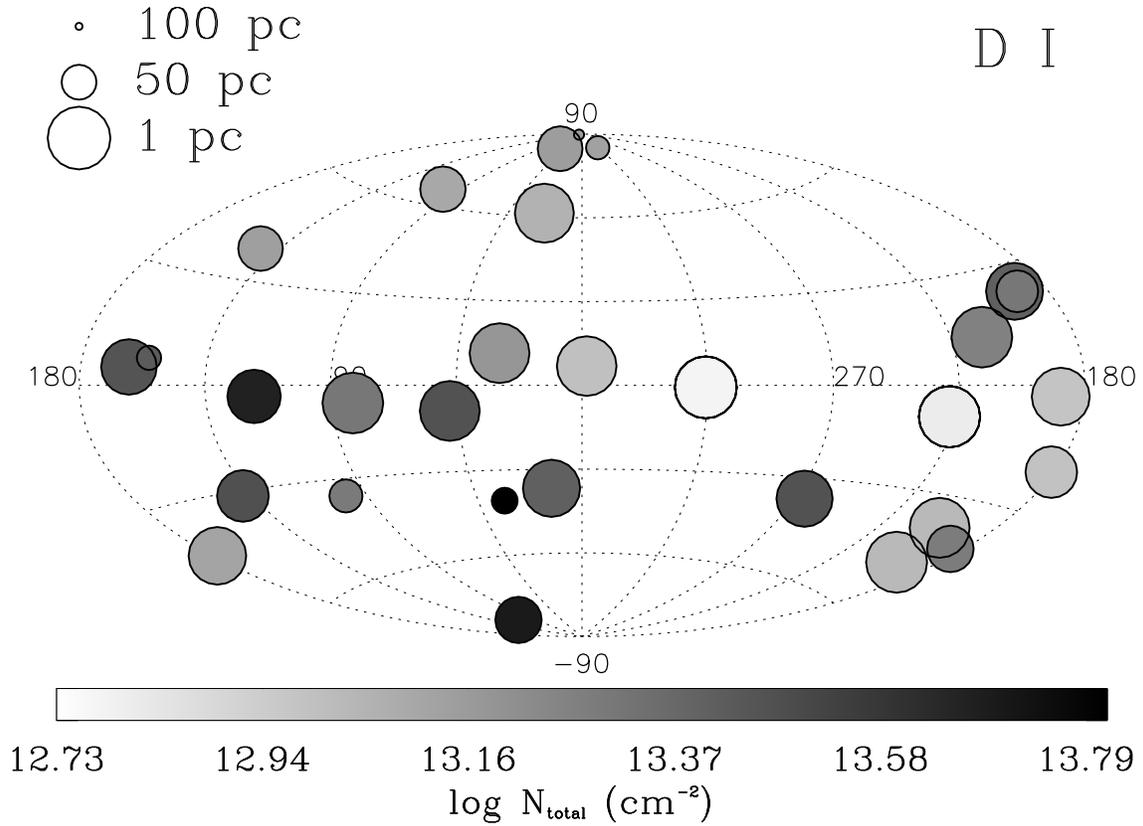}
\caption{The total column density of \ion{D}{1} along different lines of sight is shown in Galactic coordinates.  The size of each symbol is inversely proportional to the distance to the star, and the shading of the symbol indicates the total column density, as shown by the scale at the bottom of each plot.  Tables~\ref{sw_table1}, \ref{sw_table2_5}, and \ref{sw_table3} can be used to identify each particular sightline.  The total column densities are clearly spatially correlated.  The highest column densities tend to be in the southern Galactic Center direction, and the lowest column densities tend to be in the northern anti-Galactic Center direction.  \label{sw_fig12}}
\end{figure}

\begin{figure}
\epsscale{.9}
\plotone{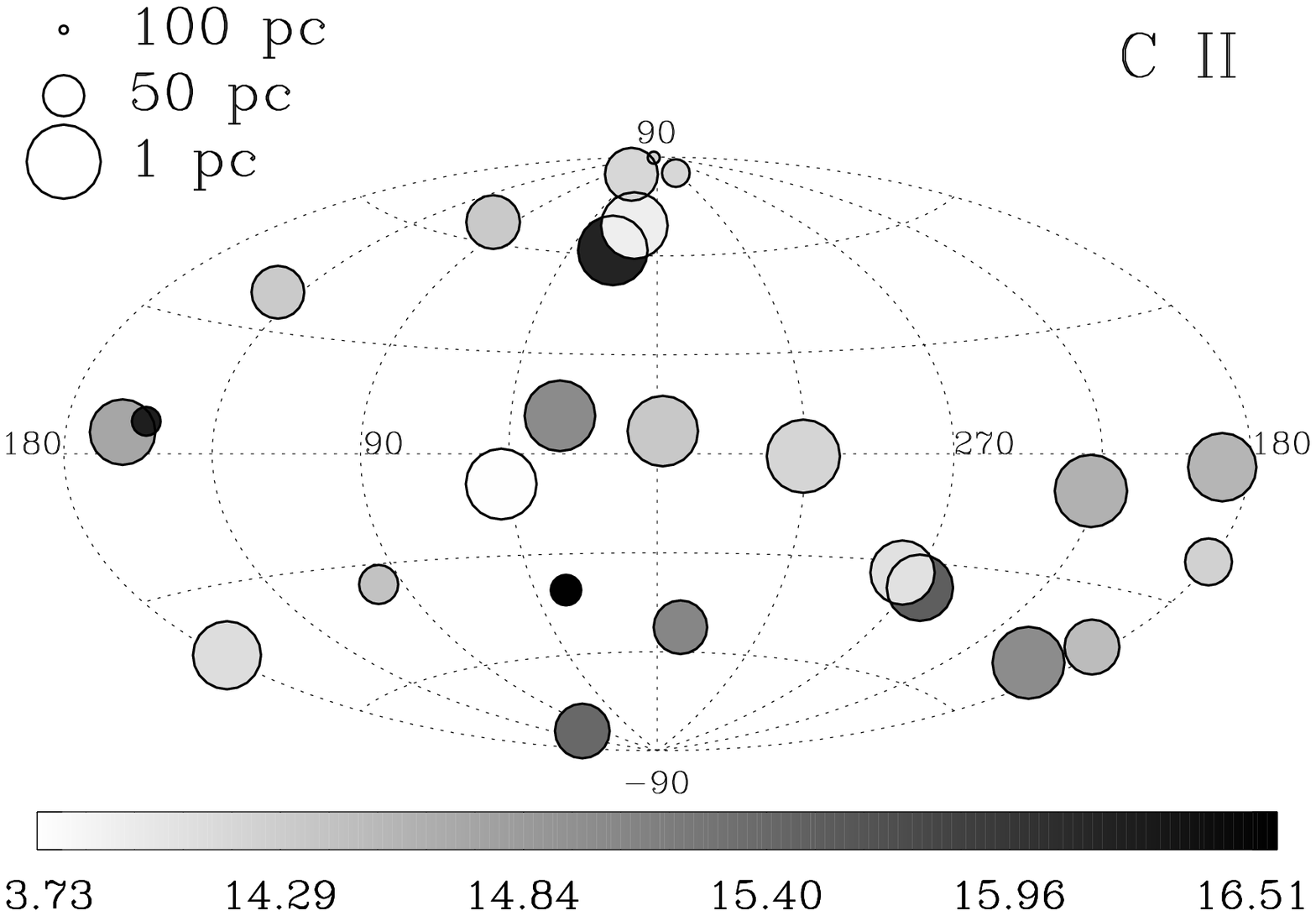}
\caption{Similar to Figure~\ref{sw_fig12}, but displaying the total column density of \ion{C}{2} in Galactic coordinates.  Tables~\ref{sw_table1}, \ref{sw_table2_5}, and \ref{sw_table4} can be used to identify each particular sightline.  \label{sw_fig13}}
\end{figure}

\begin{figure}
\epsscale{.9}
\plotone{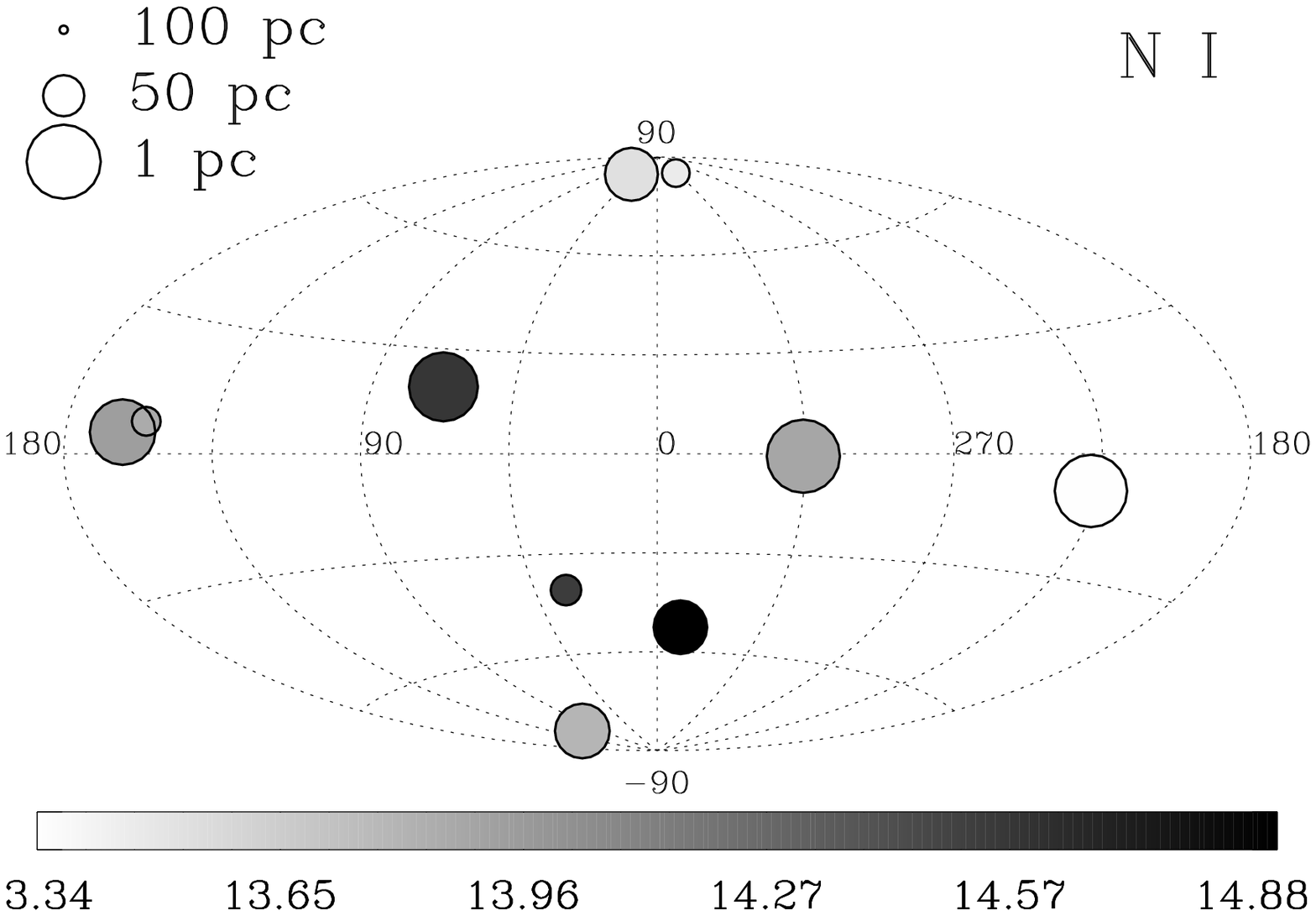}
\caption{Similar to Figure~\ref{sw_fig12}, but displaying the total column density of \ion{N}{1} in Galactic coordinates.  Tables~\ref{sw_table1}, \ref{sw_table2_5}, and \ref{sw_table5} can be used to identify each particular sightline.  \label{sw_fig14}}
\end{figure}

\begin{figure}
\epsscale{.9}
\plotone{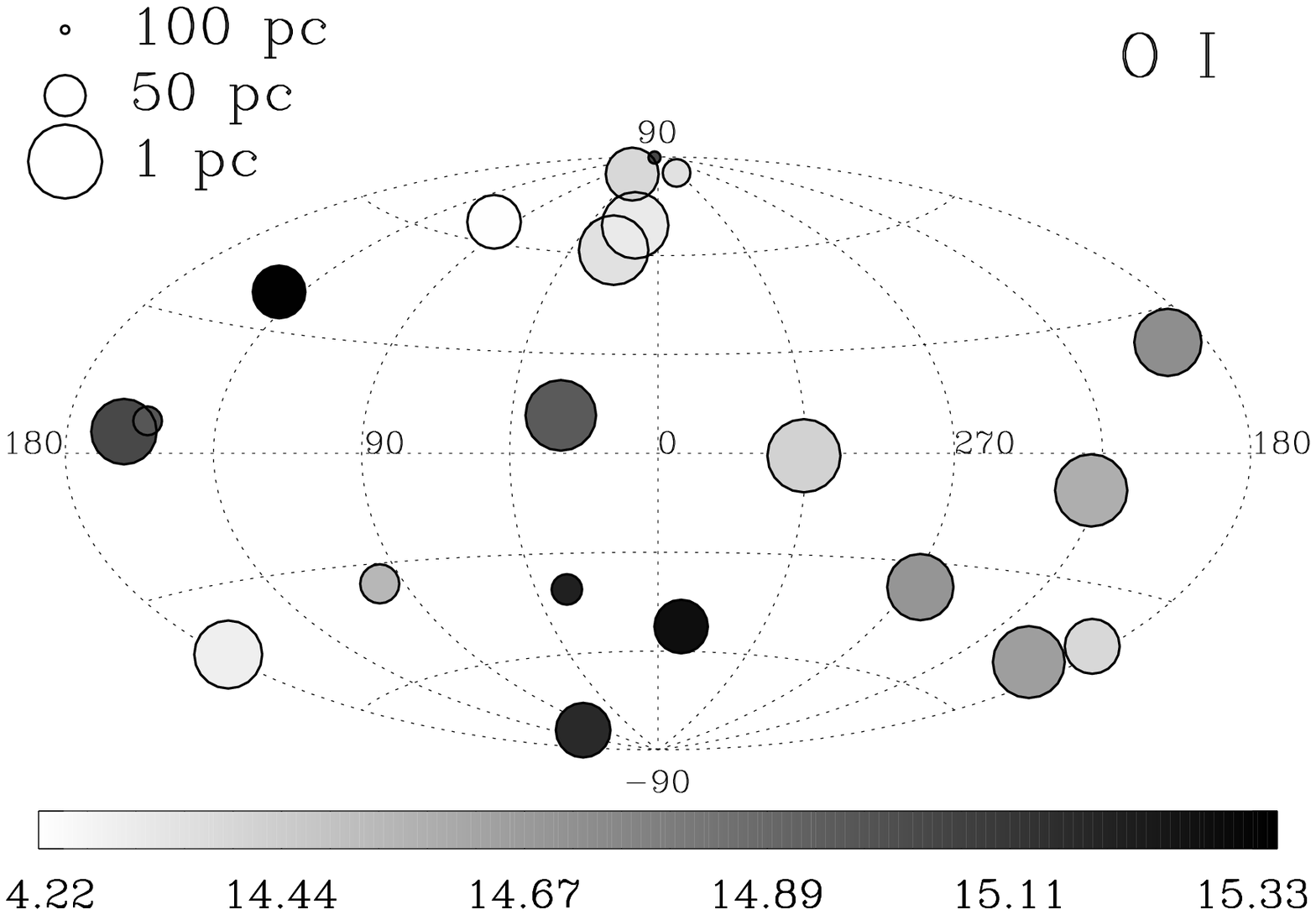}
\caption{Similar to Figure~\ref{sw_fig12}, but displaying the total column density of \ion{O}{1} in Galactic coordinates.  Tables~\ref{sw_table1}, \ref{sw_table2_5}, and \ref{sw_table6} can be used to identify each particular sightline.  \label{sw_fig15}}
\end{figure}

\begin{figure}
\epsscale{.9}
\plotone{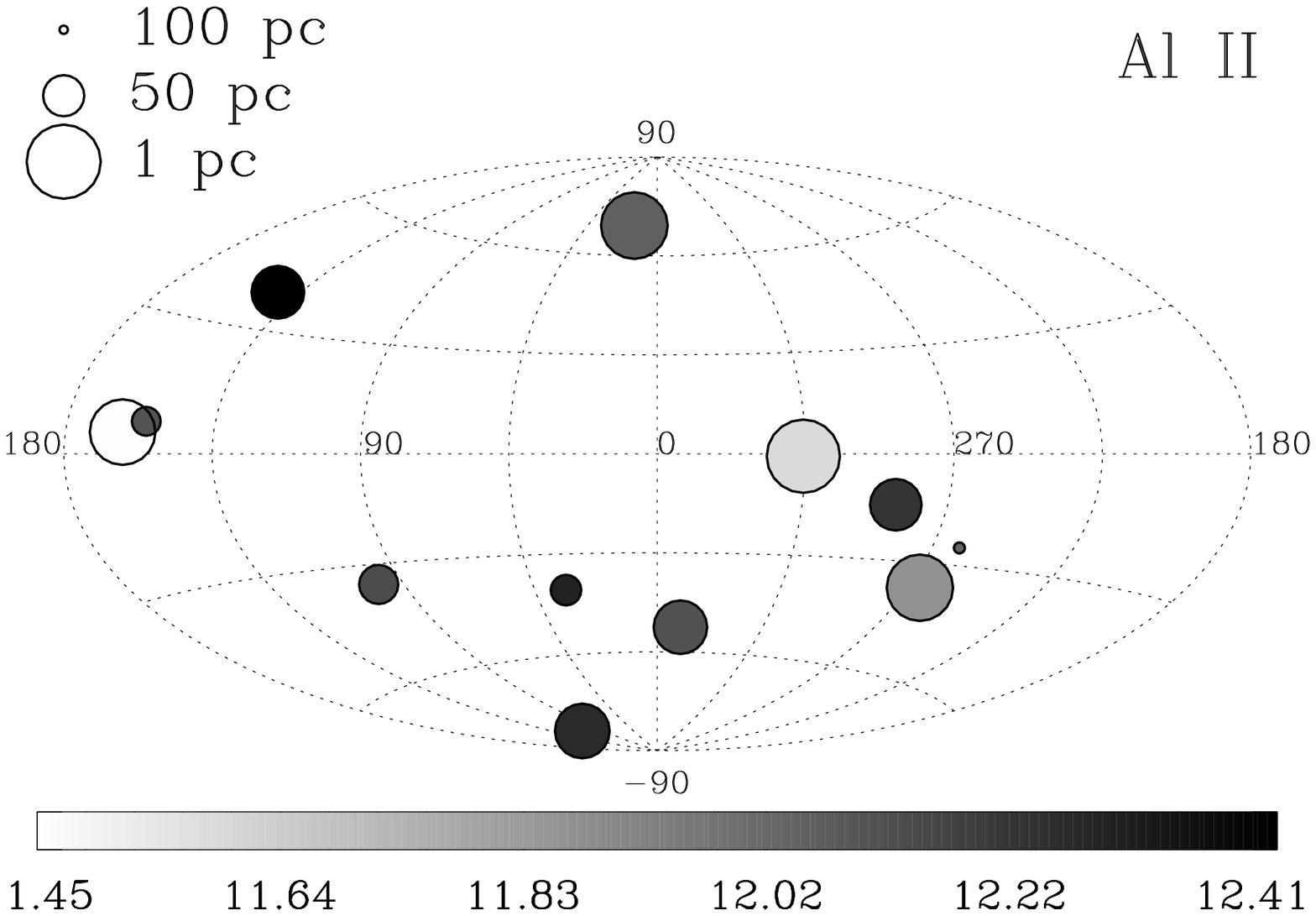}
\caption{Similar to Figure~\ref{sw_fig12}, but displaying the total column density of \ion{Al}{2} in Galactic coordinates.  Tables~\ref{sw_table1}, \ref{sw_table2_5}, and \ref{sw_table7} can be used to identify each particular sightline.  \label{sw_fig16}}
\end{figure}

\begin{figure}
\epsscale{.9}
\plotone{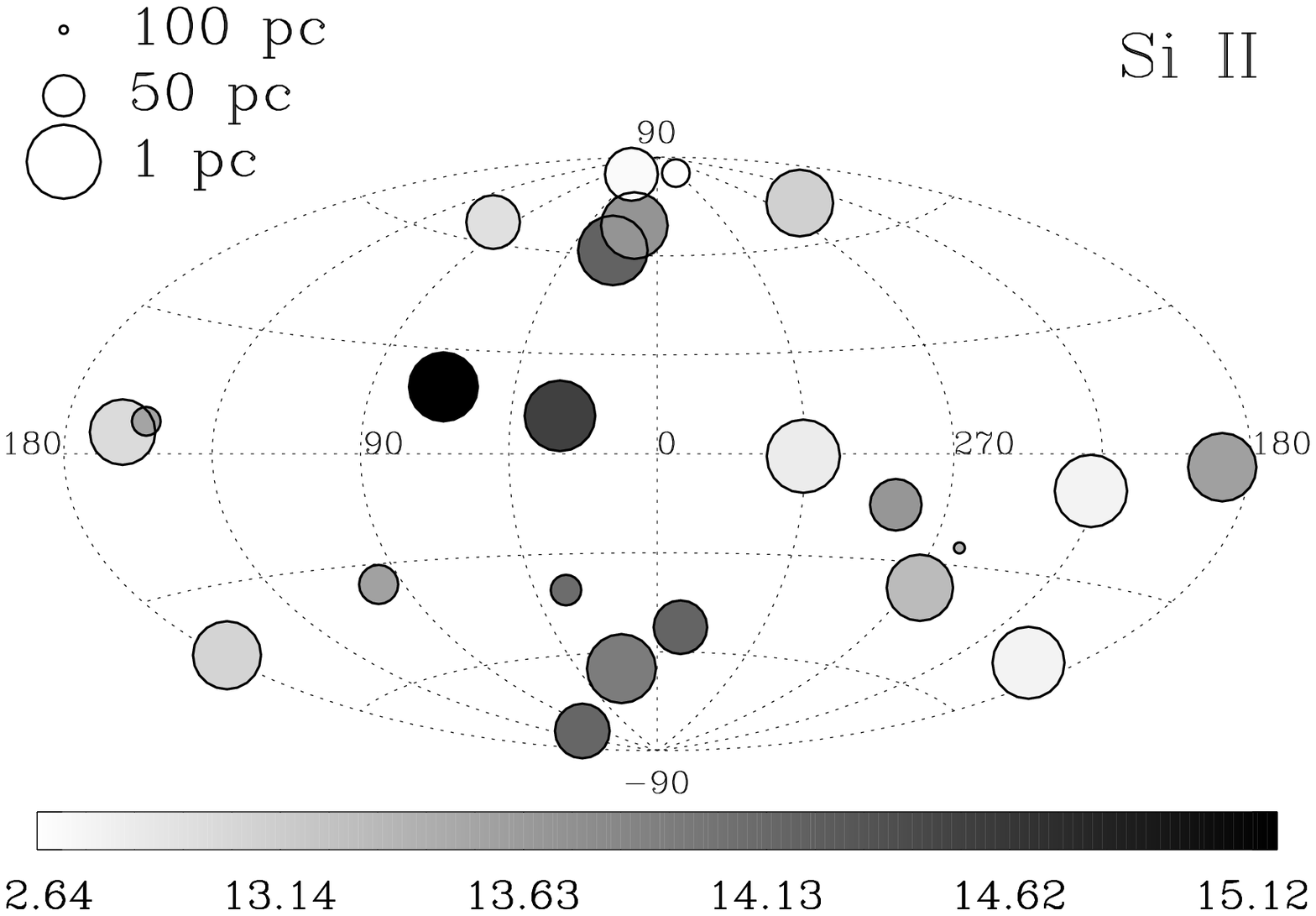}
\caption{Similar to Figure~\ref{sw_fig12}, but displaying the total column density of \ion{Si}{2} in Galactic coordinates.  Tables~\ref{sw_table1}, \ref{sw_table2_5}, and \ref{sw_table8} can be used to identify each particular sightline.  \label{sw_fig17}}
\end{figure}

\begin{figure}
\epsscale{.8}
\plotone{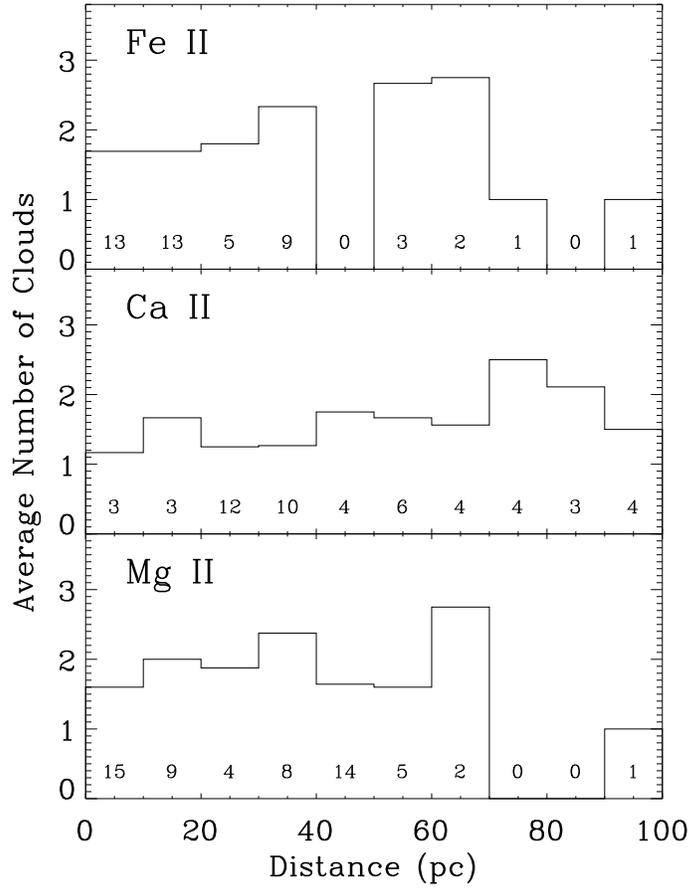}
\caption{The distribution of the average number of velocity components (clouds) along sightlines as a function of the distance to the background star.  The binsize is 10\,pc.  The total number of sightlines included in the average is given near the bottom of each bin.  Although the distance to each cloud is not known, the distance to the star provides an upper limit.  The specific morphology of the LISM can induce biases into the distribution.  For example, the sharp decline in the average number of absorbers for \ion{Fe}{2} and \ion{Mg}{2} at distances approaching 100\,pc, results from the few stars observed at these distances being located at high latitudes, where little LISM material is detected (see Figures~\ref{sw_fig12} to \ref{sw_fig17}).  The relatively constant average number of absorbers with distance indicates that the distribution of LISM clouds within the Local Bubble is not uniform but instead is concentrated close to the Sun.  
\label{sw_fig26}}
\end{figure}

Based on the number of observed LISM absorbers for each ion, given in Table~\ref{sw_table11}, the best sampling of the distribution of LISM clouds will come from \ion{Mg}{2} and \ion{Fe}{2}.  By comparing the number of velocity components (clouds) along a given sightline with the distance of the background star, we can investigate the rough distribution of warm clouds within 100\,pc.  In Figure~\ref{sw_fig26}, the average number of clouds along the line of sight toward stars located in 10\,pc bins is compared for the heavy ions, \ion{Fe}{2}, \ion{Ca}{2}, and \ion{Mg}{2}, based on the analysis of only high-resolution spectra presented here and in \citet{red02}.  It is remarkable that the average number of clouds remains nearly constant out to 100\,pc.  Only a very slight increasewith stellar distance is detected, indicating that the distribution of clouds in the LISM is not uniform, but instead is concentrated near the Sun.  The specific morphology of the LISM can induce biases into these estimates, particularly if only one star is used in a particular distance bin, as is the case for \ion{Fe}{2} and \ion{Mg}{2} at distances $>\,70$\,pc.  These particular sightlines happen to be at high latitudes, where little LISM material is detected (see Figures~\ref{sw_fig12} to \ref{sw_fig17}).  The specific morphology of the LISM will be the subject of a future paper in this series.

Previous estimates of the number density of clouds in the ISM have been between 8 and 10 cloud components per kpc \citep{spitzer78}.  These, relatively low resolution observations, were of more distant sightlines that traversed a more varied ISM environment and more clouds.  Our estimate, as shown in Figure~\ref{sw_fig26} is $\sim\,20$ clouds per kpc.  Although we only sample a relatively simple ISM environment, namely warm partially ionized clouds within the Local Bubble, our data are not severely blended because of the short lines of sight and high spectral resolution.  The cloud density estimated from more distant sightlines may suffer from such severe blending that many clouds are undetected, leading to an underestimate of the number of clouds along the line of sight.

\clearpage
\begin{figure}
\epsscale{.69}
\plotone{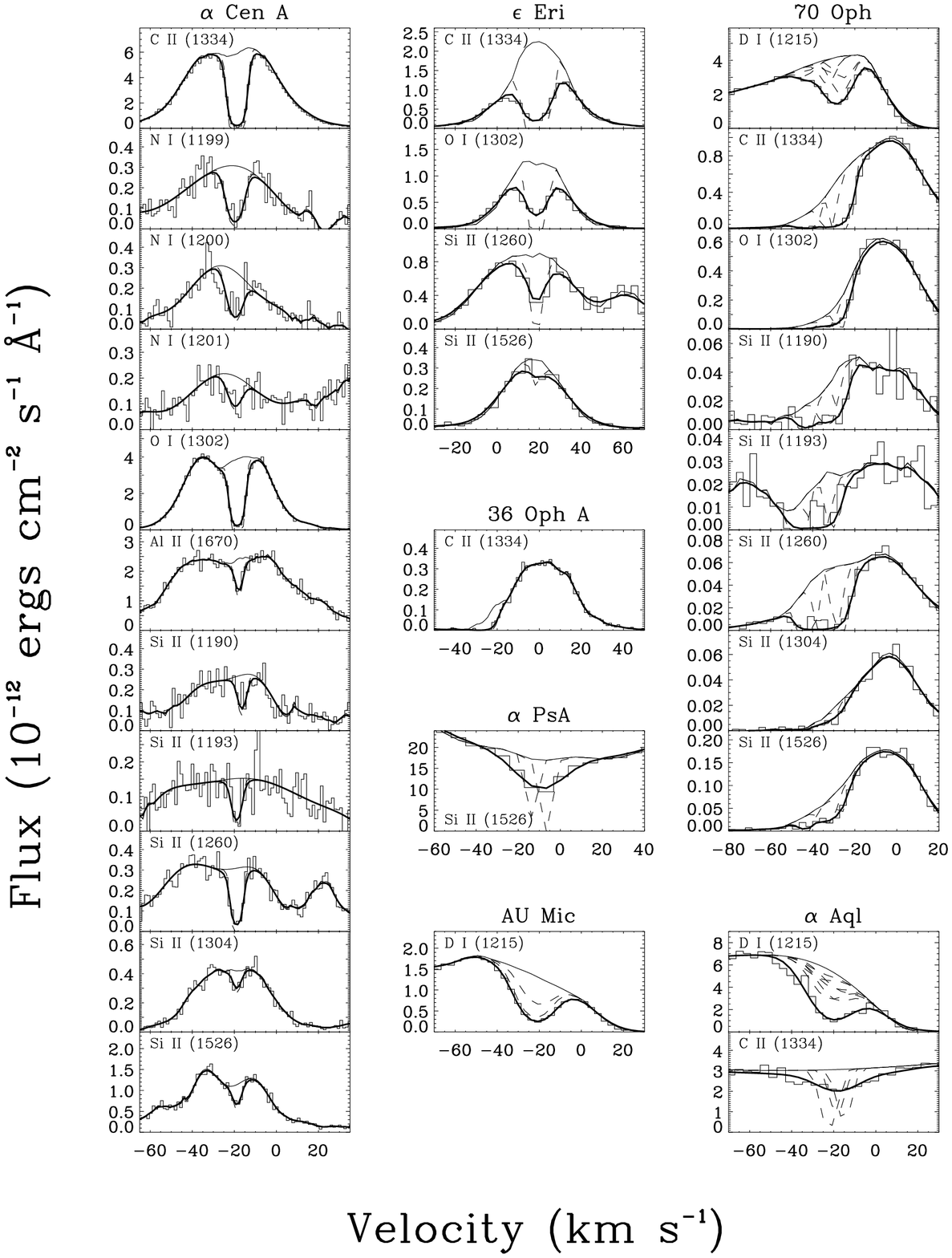}
\caption{Fits to the light ion LISM absorption lines for 32 lines of sight toward stars located within 100\,pc.  The name of the target star is given above each group of plots, and the resonance line ion (\ion{D}{1}, \ion{C}{2}, \ion{N}{1}, \ion{O}{1}, \ion{Al}{2}, or \ion{Si}{2}) and wavelength (\AA) are identified within each individual plot.  The data are shown in histogram form.  The thin solid lines are our estimates of the intrinsic stellar flux across the absorption lines.  The dashed lines are the best-fit individual absorption lines before convolution with the instrumental profile.  The thick solid line represents the combined absorption fit after convolution with the instrumental profile.  The spectra are plotted versus heliocentric velocity.  The parameters for these fits are given in Tables~\ref{sw_table3} through \ref{sw_table8}.  \label{sw_fig1}}
\end{figure}

\clearpage
\begin{figure}
\epsscale{.69}
\plotone{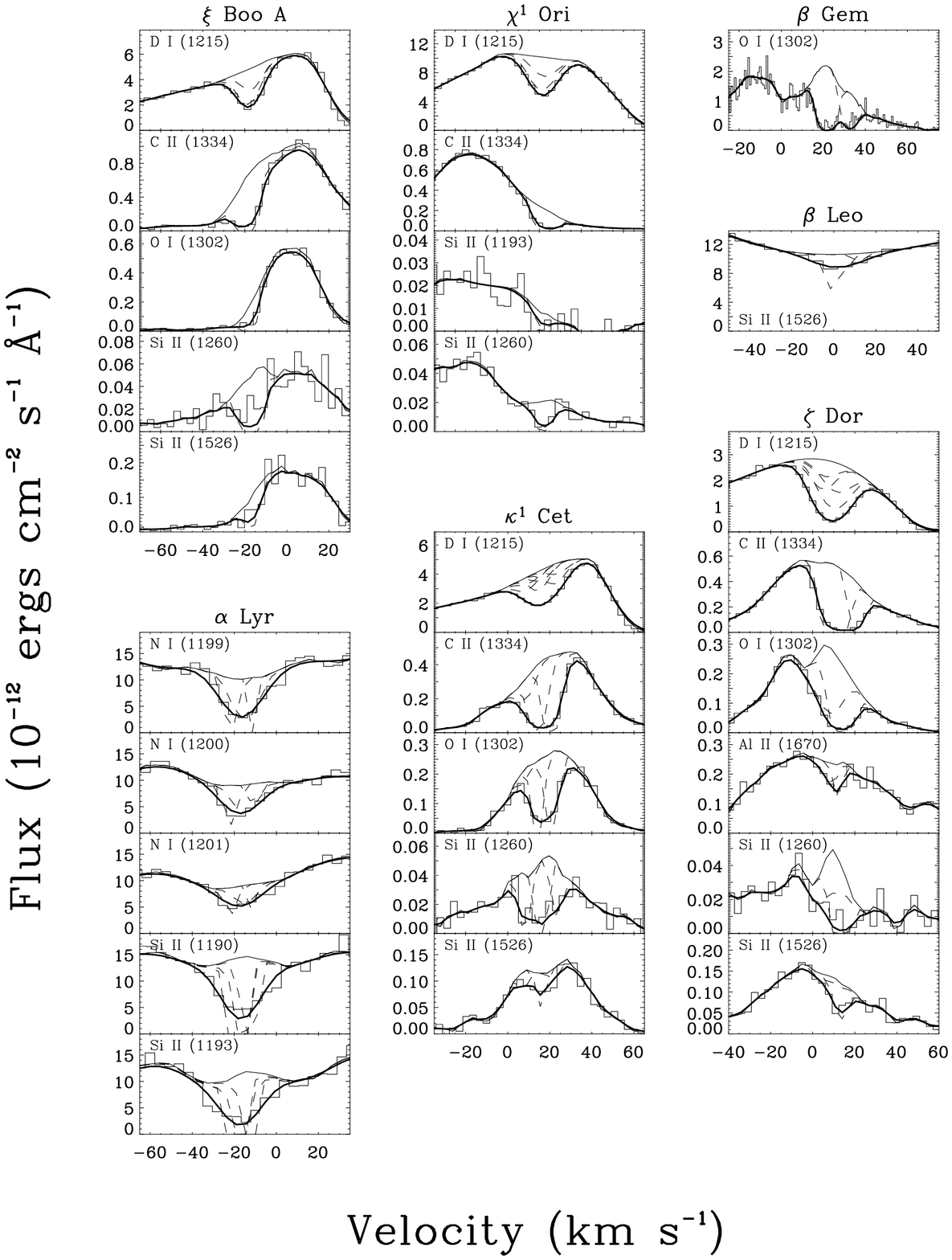}
\caption{Continuation of Figure~\ref{sw_fig1}.\label{sw_fig2}}
\end{figure}

\clearpage
\begin{figure}
\epsscale{.69}
\plotone{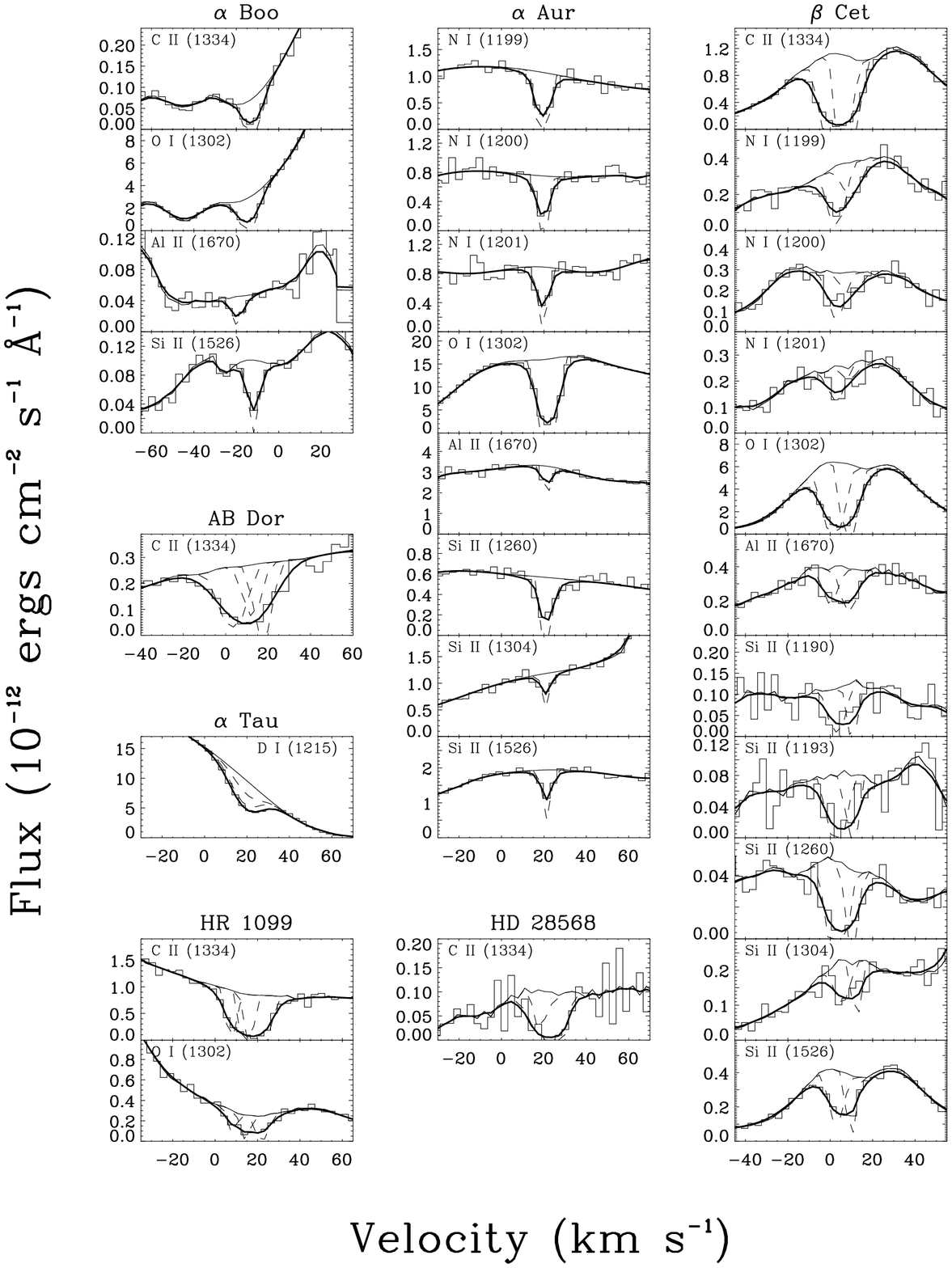}
\caption{Continuation of Figure~\ref{sw_fig1}.\label{sw_fig3}}
\end{figure}

\clearpage
\begin{figure}
\epsscale{.69}
\plotone{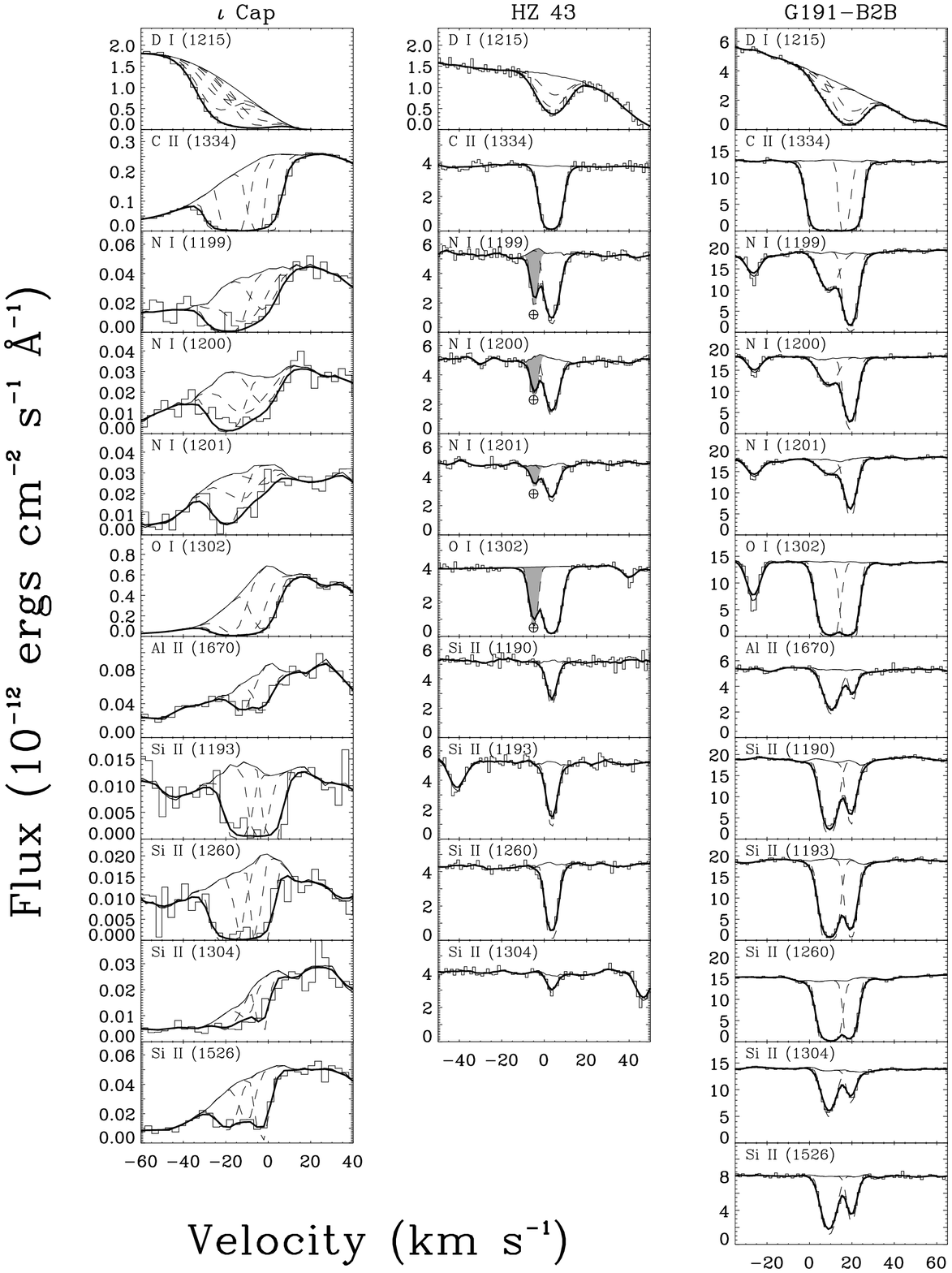}
\caption{Continuation of Figure~\ref{sw_fig1}.  The shaded absorption component are due to terrestrial airglow.\label{sw_fig4}}
\end{figure}

\clearpage
\begin{figure}
\epsscale{.69}
\plotone{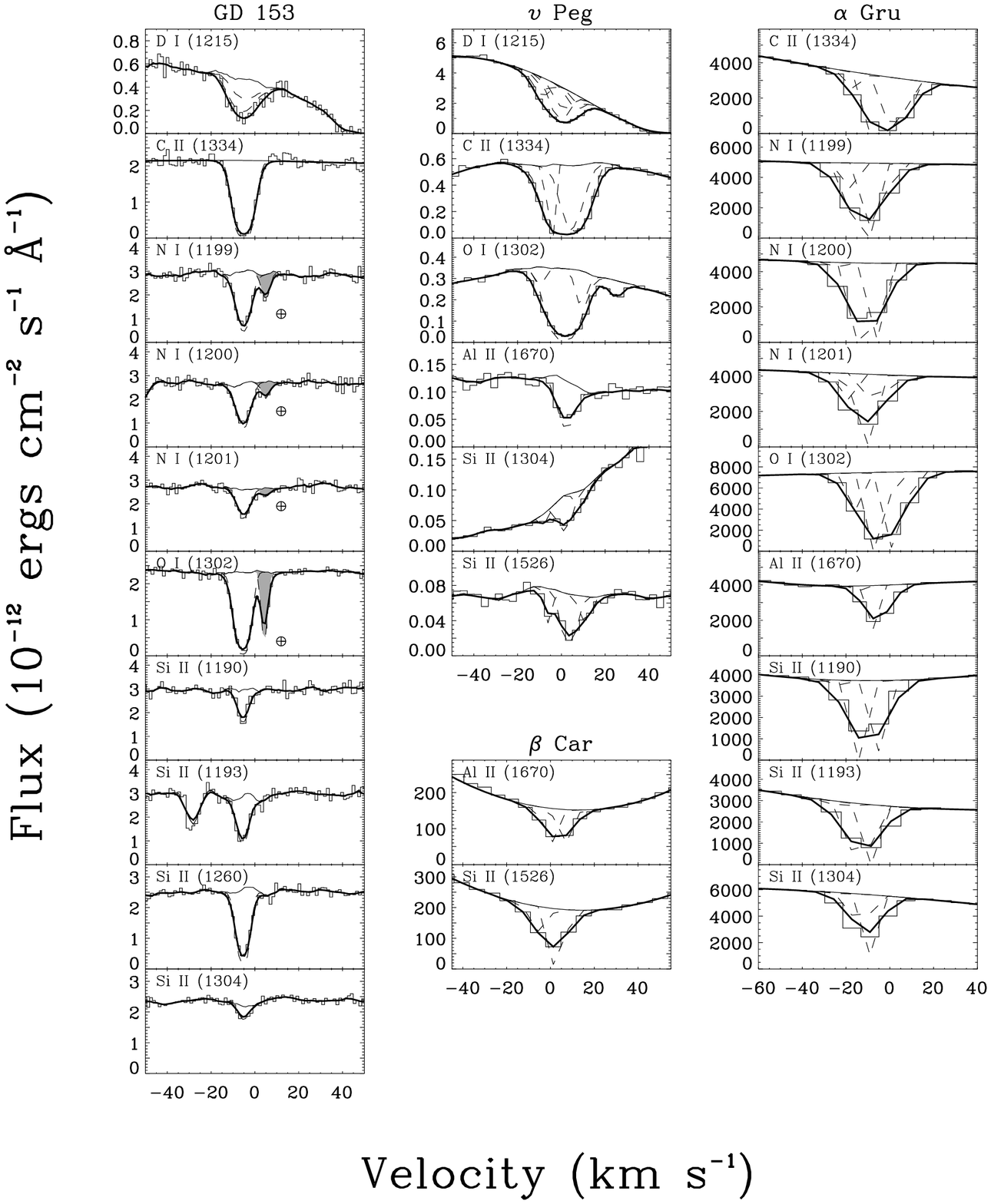}
\caption{Continuation of Figure~\ref{sw_fig1}.  The shaded absorption component are due to terrestrial airglow.\label{sw_fig5}}
\end{figure}

\clearpage
\begin{figure}
\epsscale{.69}
\plotone{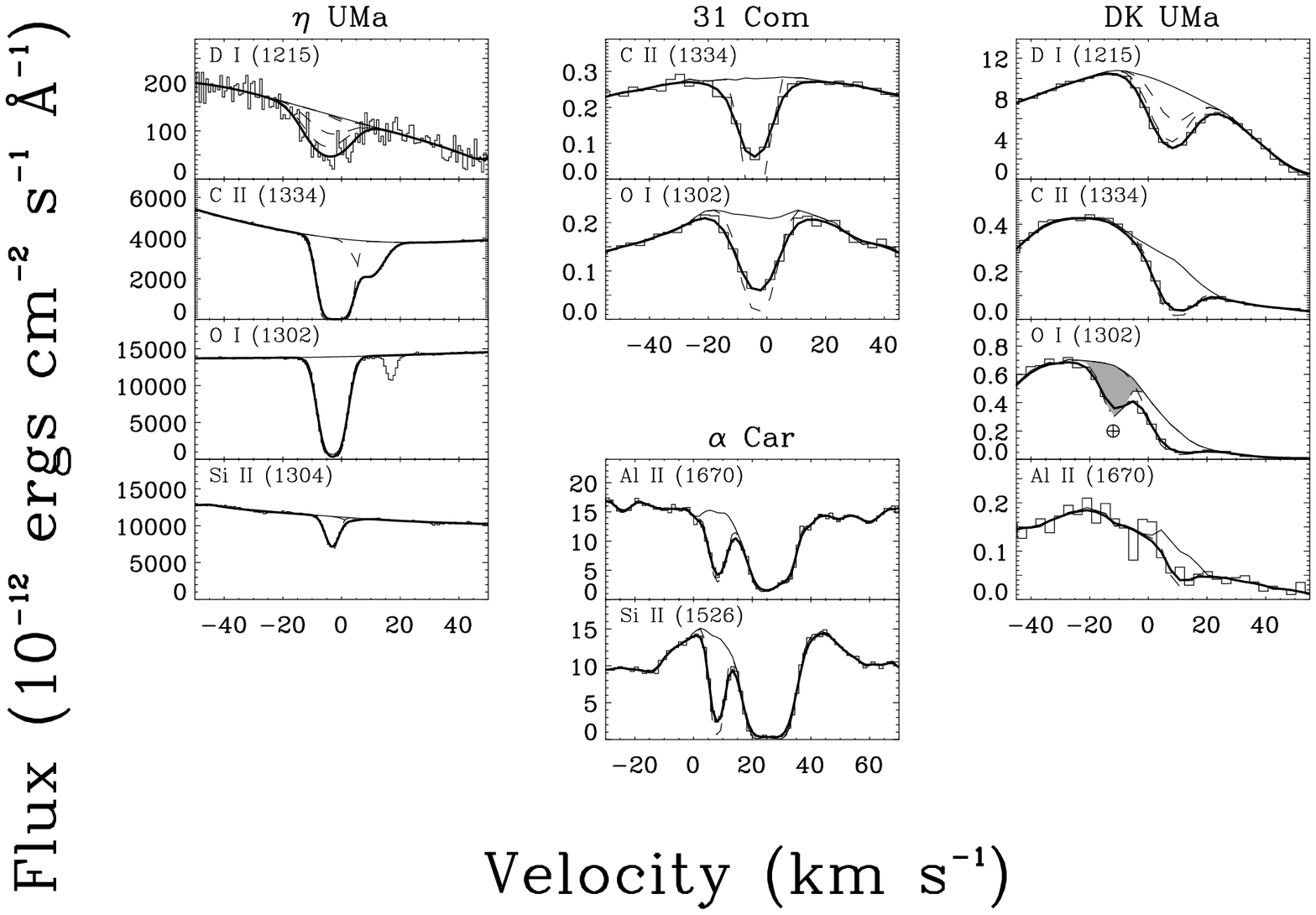}
\caption{Continuation of Figure~\ref{sw_fig1}.  The shaded absorption component is due to terrestrial airglow.\label{sw_fig6}}
\end{figure}

\clearpage

\clearpage

\clearpage

\clearpage

\clearpage

\clearpage

\clearpage

\clearpage

\clearpage

\clearpage

\clearpage

\clearpage
\begin{figure}
\epsscale{.69}
\plotone{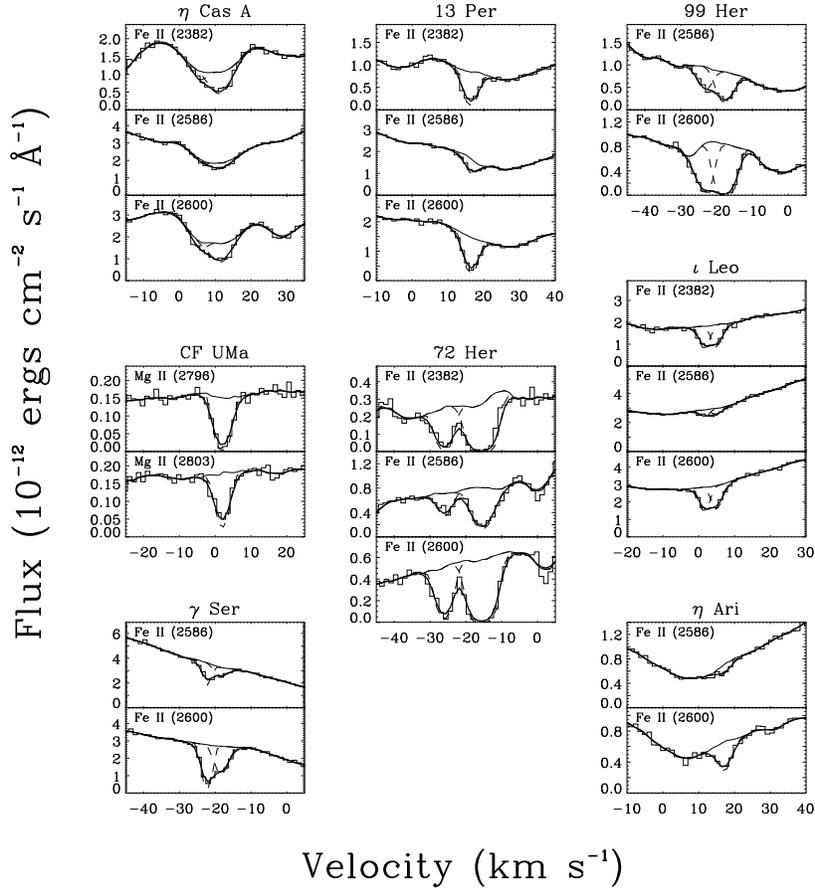}
\caption{Fits to the \ion{Mg}{2} and \ion{Fe}{2} lines for 16 lines of sight toward stars within 100\,pc.  These sightlines complement the \ion{Mg}{2} and \ion{Fe}{2} LISM absorption line sample presented by \citet{red02}.  The name of the target star is given above each group of plots, and the resonance line is identified within each individual plot.  The data are shown in histogram form.  The thin solid lines are our estimates for the intrinsic stellar flux across the absorption lines.  The dashed lines are the best-fit individual absorption lines before convolution with the instrumental profile.  The thick solid line represents the combined absorption fit after convolution with the instrumental profile.  The spectra are plotted versus heliocentric velocity.  The parameters for these fits are given in Table~\ref{sw_table9} for \ion{Mg}{2} and in Table~\ref{sw_table10} for \ion{Fe}{2}.  \label{sw_fig7}}
\end{figure}

\clearpage
\begin{figure}
\epsscale{.69}
\plotone{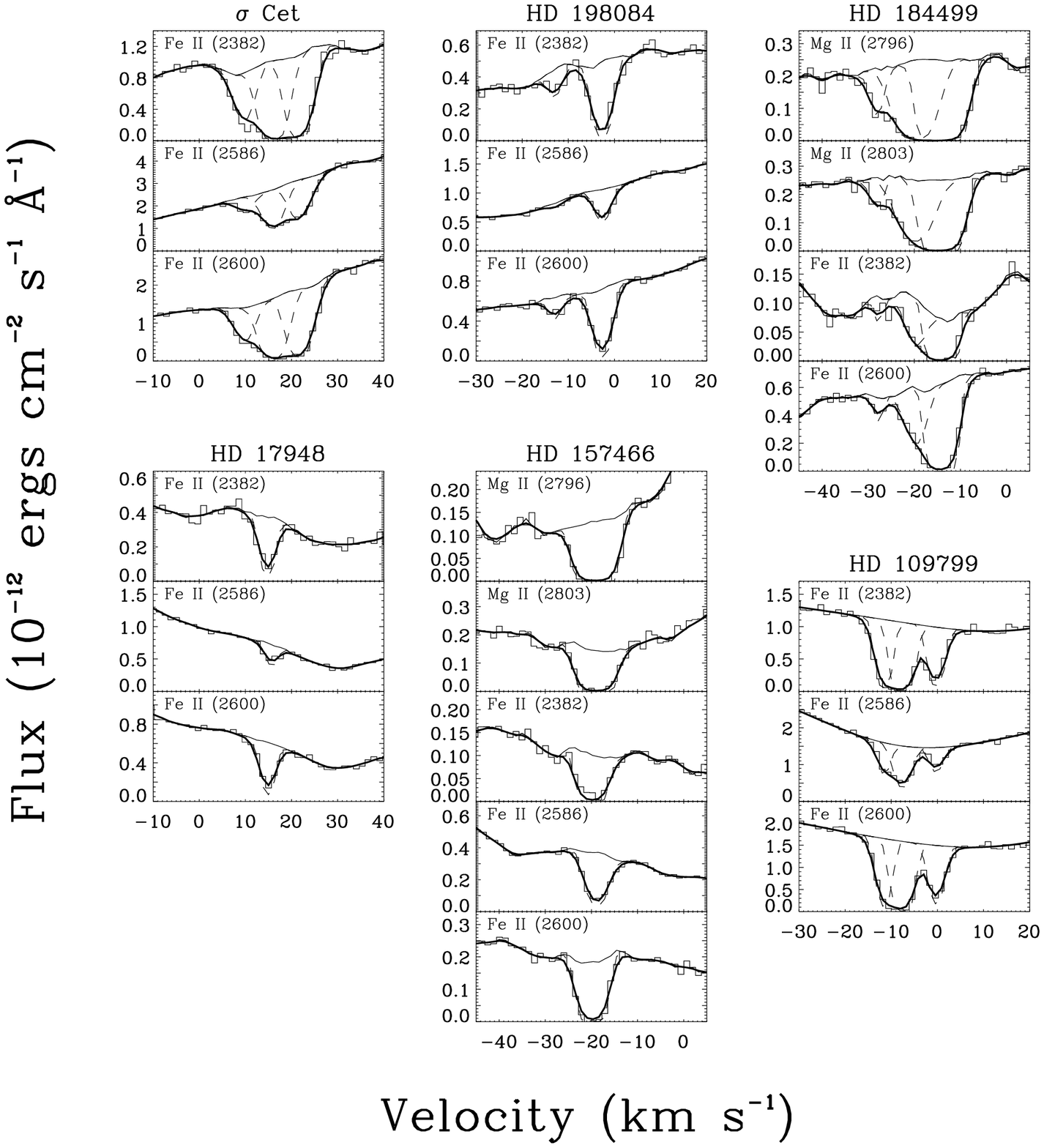}
\caption{Continuation of Figure~\ref{sw_fig7}.\label{sw_fig8}}
\end{figure}

\clearpage
\begin{figure}
\epsscale{.69}
\plotone{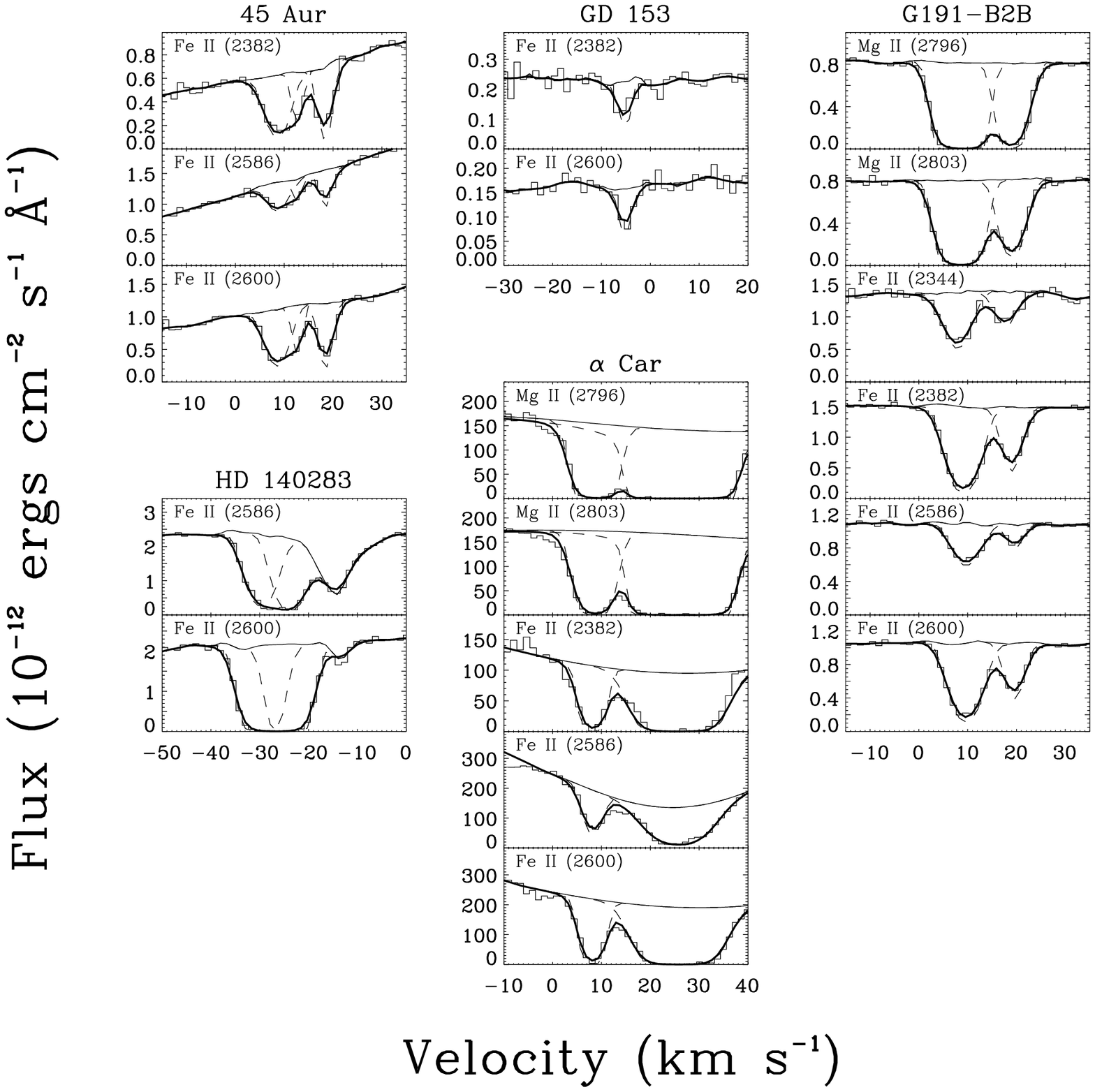}
\caption{Continuation of Figure~\ref{sw_fig7}.\label{sw_fig8b}}
\end{figure}

\clearpage

\clearpage
\begin{center}
\begin{deluxetable}{llcccccccc}
\tabletypesize{\tiny}
\tablewidth{0pt}
\tablecaption{Parameters for Stars within 100~pc with Light Ion Observations \label{sw_table1}}
\tablehead{ HD & Other & Spectral & $m_{\rm V}$ & $v_{\rm R}$ & 
$l$ & $b$ & distance \tablenotemark{a} & 
$v_{\rm{LIC}}$\tablenotemark{b} & $v_{\rm{G}}$\tablenotemark{c} \\
\# & Name & Type && (km s$^{-1}$) & ($^{\circ}$) & ($^{\circ}$) & (pc) & 
(km s$^{-1}$)  & (km s$^{-1}$)}
\startdata
128620 &$\alpha$ Cen A&G2V&--0.0\phantom{--}&--24.6&315.7&{\phn}--0.7  & {\phn}1.3 & --15.6 & --18.0\\
128621 &$\alpha$ Cen B&K1V&1.3&--20.7&315.7 & {\phn}--0.7  & {\phn}1.3 & --15.6 & --18.0\\
48915  &$\alpha$ CMa A&A1V&--1.5\phantom{--}&{\phn}--7.6& 227.2 & {\phn}--8.9  & {\phn}2.6 & \phantom{--}19.5 & \phantom{--}21.6 \\
48915B &$\alpha$ CMa B&DA&8.4&\nodata& 227.2 & {\phn}--8.9  & {\phn}2.6 & \phantom{--}19.5 & \phantom{--}21.6 \\
22049  &$\epsilon$ Eri&K2V&3.7&\phantom{--}15.5 & 195.8 & --48.1 & {\phn}3.2 & \phantom{--}21.6 & \phantom{--}25.7 \\
201091 &61 Cyg A& K5V  & 5.2  &--64.3&{\phn}82.3  & {\phn}--5.8  & {\phn}3.5 & {\phn}--5.1  & {\phn}--4.7 \\
61421  &$\alpha$ CMi&F5V& 0.3 &{\phn}--3.2& 213.7 & \phantom{--}13.0   & {\phn}3.5 & \phantom{--}19.7 & \phantom{--}21.1 \\
26965  &40 Eri A& K1V  & 4.4  &--42.2&200.8 & --38.1 & {\phn}5.0 & \phantom{--}23.2 & \phantom{--}27.2 \\
\enddata
\tablenotetext{a}{$Hipparcos$ distances \citep{perry97}.}
\tablenotetext{b}{Predicted projected LIC velocity from \citet{lall92}.}
\tablenotetext{c}{Predicted projected G velocity from \citet{lall92}.}
\tablenotetext{d}{\citet{duflot95}}
\tablenotetext{e}{\citet{brandt01}}
\tablenotetext{f}{\citet{vennes94}}
\tablenotetext{g}{\citet{reid88}}
\end{deluxetable}
\end{center}

\clearpage
\begin{center}
\begin{deluxetable}{llllccclcl}
%\rotate
\tabletypesize{\tiny}
\tablewidth{0pt}
\tablecaption{{\it HST} Observational Parameters \label{sw_table2}}
\tablehead{ HD & Other & Instrument & Grating & Spectral & Resolution & Exposure & PI & Program & Date \\
\# & Name &&& Range && Time & Name & ID & \\ 
&&&& (\AA) & ($\lambda/\Delta\lambda$) & (s) &&&}
\startdata
128620 &$\alpha$ Cen A& STIS & E140H & 1140-1335 & 114000 & 4695.0 & LINSKY & 7263 & 1999 Feb 12\\
               && STIS & E140H & 1315-1517 & 114000 & 4695.0 & LINSKY & 7263 & 1999 Feb 12\\
               && STIS & E140H & 1497-1699 & 114000 & 4695.0 & LINSKY & 7263 & 1999 Feb 12\\
48915  &$\alpha$ CMa A & GHRS & G160M & 1645-1679 &  20000 & 1740.8 & LALLEMENT & 2461 & 1992 Aug 15 \\
               && GHRS & G160M & 1651-1686 &  20000 &  108.8 & WAHLGREN & 3496 & 1993 Mar 16 \\
               && GHRS & G160M & 1651-1686 &  20000 &  108.8 & WAHLGREN & 3496 & 1993 May 8 \\
22049  &$\epsilon$ Eri& STIS & E140M & 1140-1735 &  45800 & 5798.0 & JORDAN & 7479 & 2000 Mar 17 \\
165341 &70 Oph  & STIS & E140M & 1140-1735 &  45800 & 7809.0 & AYRES & 8280 & 2000 Aug 23 \\
\enddata
\end{deluxetable}
\end{center}

\clearpage
\begin{deluxetable}{llcccccccc}
\tablewidth{0pt}
\tabletypesize{\tiny}
\tablecaption{Inventory of LISM Absorption Line Detections \label{sw_table2_5}}
\tablehead{ HD & Other & \multicolumn{6}{c}{Moderate to High Resolution Light Ions} & \multicolumn{2}{c}{High Resolution Heavy Ions} \\
\# & Name & D\,I & C\,II & N\,I & O\,I & Al\,II & Si\,II & Mg\,II & Fe\,II}
\startdata
128620 &$\alpha$ Cen A& ($\surd$) & $\surd$ & $\surd$ & $\surd$ & $\surd$ & $\surd$ & ($\surd$) & ($\surd$) \\
128621 &$\alpha$ Cen B& ($\surd$) & & & & & & ($\surd$) & ($\surd$) \\
48915  &$\alpha$ CMa A& ($\surd$) & ($\surd$) & & & $<$ & & ($\surd$) & ($\surd$) \\
48915B &$\alpha$ CMa B& ($\surd$) & ($\surd$) & ($\surd$) & ($\surd$) & & ($\surd$) & ($\surd$) & ($\surd$) \\
22049  &$\epsilon$ Eri& ($\surd$) & $\surd$ & & $\surd$ & $<$ & $\surd$ & ($\surd$) & ($\surd$) \\
201091 &61 Cyg A      & ($\surd$) & & & & & & ($\surd$) & \\
61421  &$\alpha$ CMi  & ($\surd$) & & & & & & ($\surd$) & ($\surd$) \\
26965  &40 Eri A      & ($\surd$) & & & & & & ($\surd$) & \\
\enddata
\tablecomments{The $\surd$ symbol indicates a detection of LISM absorption.  The $<$ symbol indicates no LISM absorption is detected, but an upper limit is provided.  Symbols in parentheses indicate the LISM absorption analysis has already been presented in the literature.  All detections that have not been previously analyzed are presented in this work.  For the light ions, references for analyzed sightlines can be found in Tables~\ref{sw_table3}-\ref{sw_table8}.  Details regarding the heavy ion measurements can be found in \citet{red02} and Appendix~\ref{app}.}
\end{deluxetable}

\clearpage
\begin{deluxetable}{llccccl}
\tablewidth{0pt}
\tabletypesize{\tiny}
\tablecaption{Fit Parameters for \ion{D}{1} ISM Velocity Components \label{sw_table3}}
\tablehead{ HD & Other & Component & $v$ & $b$ & $\log N_{\rm DI}$ & Reference \\
\# & Name &\# & (km s$^{-1}$) & (km s$^{-1}$) & $\log (\rm cm^{-2})$ & }
\startdata
128620 &$\alpha$ Cen A& 1 & --18.2 $\pm$ 0.1 & 6.55 $\pm$ 0.27 & 12.799 $\pm$ 0.015 & 1 \\
128621 &$\alpha$ Cen B& 1 & --18.2 $\pm$ 0.1 & 6.88 $\pm$ 0.17 & 12.775 $\pm$ 0.009 & 1 \\
48915  &$\alpha$ CMa A & 1 & 19.5 & \nodata & 12.43$^{+0.14}_{-0.20}$ & 2 \\
       &         & 2 & 13.7 & \nodata & 12.43$^{+0.14}_{-0.20}$ & 2 \\
48915B &$\alpha$ CMa B & 1 & 17.6 & 8.14 & 12.81$^{+0.09}_{-0.11}$ & 3 \\
       &         & 2 & 11.7 & 5.66 & $<$ 12.48 & 3 \\
22049  &$\epsilon$ Eri& 1 & 20.5 $\pm$ 0.7 & 8.1 $\pm$ 0.4 & 13.03 $\pm$ 0.06 & 4 \\
201091 &61 Cyg A& 1 & --3.0 & 7.8 & 13.01 & 5 \\
\enddata
\tablerefs{(1) Linsky \& Wood 1996; (2) Bertin et al. 1995; (3) H\'{e}brard et al. 1999; (4) Dring et al. 1997; (5) Wood \& Linsky 1998; (6) Linsky et al. 1995; (7) this paper; (8) Landsman \& Simon 1996; (9) Wood, Linsky, \& Zank 2000; (10) Linsky et al. 1993; (11) Piskunov et al. 1997.}
\end{deluxetable}  

\clearpage
\begin{deluxetable}{llccccl}
\tablewidth{0pt}
\tabletypesize{\tiny}
\tablecaption{Fit Parameters for \ion{C}{2} ISM Velocity Components \label{sw_table4}}
\tablehead{ HD & Other & Component & $v$ & $b$ & $\log N_{\rm CII}$ & Reference \\
\# & Name &\# & (km s$^{-1}$) & (km s$^{-1}$) & $\log (\rm cm^{-2})$ & }
\startdata
128620 &$\alpha$ Cen A& 1 & --18.70 $\pm$ 0.11 & 2.93 $\pm$ 0.15 & 14.30 $\pm$ 0.12 & 1 \\
48915  &$\alpha$ CMa A & 1 & 17.6 $\pm$ 1.5 & \nodata & 14.62$^{+0.23}_{-0.28}$ & 2 \\
       &         & 2 & 11.7 $\pm$ 1.5 & \nodata & 13.78$^{+0.15}_{-0.12}$ & 2 \\
48915B &$\alpha$ CMa B & 1 & 17.6 $\pm$ 1.5 & \nodata & 14.62$^{+0.23}_{-0.28}$ & 2 \\
       &         & 2 & 11.7 $\pm$ 1.5 & \nodata & 13.78$^{+0.15}_{-0.12}$ & 2 \\
%48915  &$\alpha$ CMa A & 1 & 17.6 $\pm$ 1.5 & 3.37$^{+0.11}_{-0.21}$ & 14.62$^{+0.23}_{-0.28}$ & 2 \\
%       &         & 2 & 11.7 $\pm$ 1.5 & 3.38$^{+0.47}_{-0.31}$ & 13.78$^{+0.15}_{-0.12}$ & 2 \\
%48915B &$\alpha$ CMa B & 1 & 17.6 $\pm$ 1.5 & 3.37$^{+0.11}_{-0.21}$ & 14.62$^{+0.23}_{-0.28}$ & 2 \\
%       &         & 2 & 11.7 $\pm$ 1.5 & 3.38$^{+0.47}_{-0.31}$ & 13.78$^{+0.15}_{-0.12}$ & 2 \\
22049  &$\epsilon$ Eri& 1 & 18.50 $\pm$ 0.06 & 4.30 $\pm$ 1.06 & 15.08 $\pm$ 0.63 & 1 \\
165341 &70 Oph  & 1 & --26.5 & 3.93 $\pm$ 0.92 & 14.35 $\pm$ 0.31 & 1 \\
       &        & 2 & --32.9 & 3.40 $\pm$ 1.26 & 13.92 $\pm$ 0.26 & 1 \\
\enddata
\tablerefs{(1) this paper; (2) H\'{e}brard et al. 1999; (3) Wood \& Linsky 1997; (4) Gnaci\'{n}ski 2000.}
\end{deluxetable}  

\clearpage
\begin{deluxetable}{llccccl}
\tablewidth{0pt}
\tabletypesize{\tiny}
\tablecaption{Fit Parameters for \ion{N}{1} ISM Velocity Components\label{sw_table5}}
\tablehead{ HD & Other & Component & $v$ & $b$ & $\log N_{\rm NI}$ & Reference \\
\# & Name &\# & (km s$^{-1}$) & (km s$^{-1}$) & $\log (\rm cm^{-2})$ & }
\startdata
128620 &$\alpha$ Cen A& 1 & --19.58 $\pm$ 0.64 & 3.44 $\pm$ 1.17 & 13.89 $\pm$ 0.20 & 1 \\
48915B &$\alpha$ CMa B & 1 & 17.6 $\pm$ 1.5 & \nodata & 13.11$^{+0.04}_{-0.07}$ & 2 \\
       &         & 2 & 11.7 $\pm$ 1.5 & \nodata & 12.96$^{+0.09}_{-0.05}$ & 2 \\
%48915B &$\alpha$ CMa B & 1 & 17.6 $\pm$ 1.5 & 3.12$^{+0.11}_{-0.20}$ & 13.11$^{+0.04}_{-0.07}$ & 2 \\
%       &         & 2 & 11.7 $\pm$ 1.5 & 3.29$^{+0.44}_{-0.30}$ & 12.96$^{+0.09}_{-0.05}$ & 2 \\
172167 &$\alpha$ Lyr    & 1 & --12.06 $\pm$ 2.16 & 3.10 $\pm$ 1.87 & 13.80 $\pm$ 0.13 & 1 \\
       &        & 2 & --18.52 $\pm$ 2.77 & 3.88 $\pm$ 1.95 & 13.64 $\pm$ 0.11 & 1 \\
       &        & 3 & --20.87 $\pm$ 2.05 & 3.11 $\pm$ 1.30 & 14.45 $\pm$ 0.58 & 1 \\
34029  &$\alpha$ Aur & 1 & 19.31 $\pm$ 0.35 & 2.73 $\pm$ 0.64 & 13.94 $\pm$ 0.20 & 1 and 3 \\
4128   &$\beta$ Cet& 1 & 10.20 $\pm$ 1.73 & 3.49 $\pm$ 1.29 & 13.37 $\pm$ 0.20 & 1 \\
\enddata
\tablerefs{(1) this paper; (2) H\'{e}brard et al. 1999; (3) Wood et al. 2002; (4) Vidal-Madjar et al. 1998.}
\end{deluxetable}  

\clearpage
\begin{deluxetable}{llccccl}
\tablewidth{0pt}
\tabletypesize{\tiny}
\tablecaption{Fit Parameters for \ion{O}{1} ISM Velocity Components\label{sw_table6}}
\tablehead{ HD & Other & Component & $v$ & $b$ & $\log N_{\rm OI}$ & Reference \\
\# & Name &\# & (km s$^{-1}$) & (km s$^{-1}$) & $\log (\rm cm^{-2})$ & }
\startdata
128620 &$\alpha$ Cen A& 1 & --18.63 $\pm$ 0.03 & 2.64 $\pm$ 0.19 & 14.43 $\pm$ 0.12 & 1 \\
48915B &$\alpha$ CMa B & 1 & 17.6 $\pm$ 1.5 & \nodata & 14.53$^{+0.16}_{-0.15}$ & 2 \\
       &         & 2 & 11.7 $\pm$ 1.5 & \nodata & 13.70$^{+0.11}_{-0.10}$ & 2 \\
%48915B &$\alpha$ CMa B & 1 & 17.6 $\pm$ 1.5 & 2.93$^{+0.10}_{-0.18}$ & 14.53$^{+0.16}_{-0.15}$ & 2 \\
%       &         & 2 & 11.7 $\pm$ 1.5 & 3.23$^{+0.41}_{-0.30}$ & 13.70$^{+0.11}_{-0.10}$ & 2 \\
22049  &$\epsilon$ Eri& 1 & 18.75 $\pm$ 0.15 & 3.87 $\pm$ 0.91 & 14.66 $\pm$ 0.52 & 1 \\
165341 &70 Oph  & 1 & --26.5 & 3.32 $\pm$ 0.75 & 14.02 $\pm$ 0.08 & 1 \\
       &        & 2 & --32.9 & 2.87 $\pm$ 1.51 & 13.95 $\pm$ 0.16 & 1 \\
       &        & 3 & --43.3 & 4.76 $\pm$ 3.65 & 14.86 $\pm$ 0.31 & 1 \\
131156A&$\xi$ Boo A& 1 & --18.00 $\pm$ 0.61 & 3.12 $\pm$ 0.73 & 14.36 $\pm$ 0.14 & 1 \\
\enddata
\tablerefs{(1) this paper; (2) H\'{e}brard et al. 1999; (3) Wood et al. 2002; (4) Vidal-Madjar et al. 1998.}
\end{deluxetable}  

\clearpage
\begin{deluxetable}{llccccl}
\tablewidth{0pt}
\tabletypesize{\tiny}
\tablecaption{Fit Parameters for \ion{Al}{2} ISM Velocity Components\label{sw_table7}}
\tablehead{ HD & Other & Component & $v$ & $b$ & $\log N_{\rm AlII}$ & Reference \\
\# & Name &\# & (km s$^{-1}$) & (km s$^{-1}$) & $\log (\rm cm^{-2})$ & }
\startdata
128620 &$\alpha$ Cen A& 1 & --17.81 $\pm$ 0.21 & 2.08 $\pm$ 0.34 & 11.59 $\pm$ 0.09 & 1 \\
48915  &$\alpha$ CMa A & & \nodata & \nodata & $<$ 11.8 & 1 \\
22049  &$\epsilon$ Eri& & \nodata & \nodata & $<$ 11.3 & 1 \\
165341 &70 Oph & & \nodata & \nodata & $<$ 12.0 & 1 \\
131156A& $\xi$ Boo A  & & \nodata & \nodata & $<$ 12.1 & 1 \\
216956 &$\alpha$ PsA & & \nodata & \nodata & $<$ 12.6 & 1 \\
39587  &$\chi^1$ Ori& & \nodata & \nodata & $<$ 11.8 & 1 \\
20630  &$\kappa^1$ Cet& & \nodata & \nodata & $<$ 11.7 & 1 \\
\enddata
\tablerefs{(1) this paper; (2) Wood et al. 2002.}
\end{deluxetable}  

\clearpage
\begin{deluxetable}{llccccl}
\tablewidth{0pt}
\tabletypesize{\tiny}
\tablecaption{Fit Parameters for \ion{Si}{2} ISM Velocity Components\label{sw_table8}}
\tablehead{ HD & Other & Component & $v$ & $b$ & $\log N_{\rm SiII}$ & Reference \\
\# & Name &\# & (km s$^{-1}$) & (km s$^{-1}$) & $\log (\rm cm^{-2})$ & }
\startdata
128620 &$\alpha$ Cen A& 1 & --17.92 $\pm$ 1.28 & 2.15 $\pm$ 0.60 & 12.83 $\pm$ 0.15 & 1 \\
48915B &$\alpha$ CMa B & 1 & 17.6 $\pm$ 1.5 & \nodata & 12.48$^{+0.12}_{-0.08}$ & 2 \\
       &         & 2 & 11.7 $\pm$ 1.5 & \nodata & 12.43$^{+0.14}_{-0.09}$ & 2 \\
%48915B &$\alpha$ CMa B & 1 & 17.6 $\pm$ 1.5 & 2.23$^{+0.09}_{-0.15}$ & 12.48$^{+0.12}_{-0.08}$ & 2 \\
%       &         & 2 & 11.7 $\pm$ 1.5 & 3.01$^{+0.33}_{-0.29}$ & 12.43$^{+0.14}_{-0.09}$ & 2 \\
22049  &$\epsilon$ Eri& 1 & 18.89 $\pm$ 0.62 & 3.07 $\pm$ 0.46 & 12.76 $\pm$ 0.14 & 1 \\
165341 &70 Oph  & 1 & --26.8 & 3.01 $\pm$ 0.48 & 13.16 $\pm$ 0.13 & 1 \\
       &        & 2 & --33.2 & 2.80 $\pm$ 1.31 & 13.88 $\pm$ 0.57 & 1 \\
       &        & 3 & --43.6 & 2.65 $\pm$ 0.56 & 14.42 $\pm$ 0.37 & 1 \\
131156A&$\xi$ Boo A& 1 & --17.62 $\pm$ 0.98 & 3.97 $\pm$ 2.39 & 14.21 $\pm$ 0.32 & 1 \\
\enddata
\tablerefs{(1) this paper; (2) H\'{e}brard et al. 1999; (3) Wood et al. 2002; (4) Vidal-Madjar et al. 1998.}
\end{deluxetable}  

\clearpage
\begin{deluxetable}{lcccccc}
\tablewidth{0pt}
\tabletypesize{\tiny}
\tablecaption{Weighted Mean of LISM Parameters in the Aggregate \label{sw_table11}}
\tablehead{ & \multicolumn{3}{c}{\underline{Doppler Width Parameter $b$ (km s$^{-1}$)}} & \multicolumn{3}{c}{\underline{Column Density $\log\,N$ (cm$^{-2}$)}} \vspace{1mm}\\
Ion & Weighted & Standard & N & Weighted & Standard & N \\
&  Mean & Deviation & & Mean & Deviation & }
\startdata
D\,I   & 7.65 & 0.67 & 48 & 13.01 & 0.25 & 54 \\
C\,II  & 3.64 & 0.81 & 45 & 13.82 & 0.60 & 49 \\
N\,I   & 3.28 & 0.30 & 17 & 13.51 & 0.24 & 19 \\
O\,I   & 3.24 & 0.70 & 36 & 14.00 & 0.44 & 38 \\
Mg\,II & 2.79 & 0.55 & 107\phn & 12.45 & 0.33 & 107\phn \\
Al\,II & 2.80 & 0.60 & 19 & 11.93 & 0.17 & 19 \\
Si\,II & 2.70 & 0.34 & 41 & 13.08 & 0.34 & 43 \\
Fe\,II & 2.13 & 0.25 & 85 & 12.45 & 0.24 & 91 \\
\enddata
\end{deluxetable}

\clearpage
\begin{center}
\begin{deluxetable}{llcccccccc}
\tabletypesize{\tiny}
\tablewidth{0pt}
\tablecaption{Parameters for Stars within 100~pc with Heavy Ion Observations\tablenotemark{a} \label{sw_table_a1}}
\tablehead{ HD & Other & Spectral & $m_{\rm V}$ & $v_{\rm R}$ & 
$l$ & $b$ & distance \tablenotemark{b} & 
$v_{\rm{LIC}}$\tablenotemark{c} & $v_{\rm{G}}$\tablenotemark{d} \\
\# & Name & Type && (km s$^{-1}$) & ($^{\circ}$) & ($^{\circ}$) & (pc) & 
(km s$^{-1}$)  & (km s$^{-1}$)}
\startdata
4614 & $\eta$ Cas A&G0V& 3.5  &     & 122.6 & --5.1  & 6.0 & 11.6   & 13.8 \\
103095 &CF UMa & G8Vp  & 6.5  &--98.0&168.5 & 73.8   & 9.2 & --0.4 & --2.5 \\
142806 &$\gamma$ Ser&F6IV&3.9 & 6.7 & 27.7  & 45.7   & 11.1 & --21.2 & --25.0 \\
16895  &13 Per & F7V  & 4.1  &25.0 & 141.2 & --9.6  & 11.2 & 18.4 & 21.5 \\
157214 &72 Her & G0V &   5.4  &--78.5&55.9  & 32.3   & 14.4 &--17.3&--20.0\\
165908 &99 Her  & F7V  & 5.1  & 1.0 & 57.0  & 22.3   & 15.7 & --17.1 &--19.4\\
99028  &$\iota$ Leo&F4IV& 4.0 &--10.3&247.6 & 63.6   & 24.2 & --1.3&--3.7\\
15798  &$\sigma$ Cet&F5V&4.7  &--29.4&191.1 & --63.8 & 25.8 & 17.4 & 21.3\\
\enddata
\tablenotetext{a}{A supplemental table to Table~1 in \citet{red02}.}
\tablenotetext{b}{$Hipparcos$ distances \citep{perry97}.}
\tablenotetext{c}{Predicted projected LIC velocity from \citet{lall92}.}
\tablenotetext{d}{Predicted projected G velocity from \citet{lall92}.}
\tablenotetext{e}{\citet{reid88}}
\end{deluxetable}
\end{center}

\clearpage
\begin{center}
\begin{deluxetable}{llllccclcl}
%\rotate
\tabletypesize{\tiny}
\tablewidth{0pt}
\tablecaption{{\it HST} Observational Parameters\tablenotemark{a} \label{sw_table_a2}}
\tablehead{ HD & Other & Instrument & Grating & Spectral & Resolution & Exposure & PI & Program & Date \\
\# & Name &&& Range && Time & Name & ID & \\ 
&&&& (\AA) & ($\lambda/\Delta\lambda$) & (s) &&&}
\startdata
4614   & $\eta$ Cas A&STIS&E230H&2374-2646 & 114000 & 2060.0 & BOESGAARD & 8770 & 2001 Feb 14 \\
103095 &CF UMa  & STIS & E230H & 2624-2895 & 114000 & 2904.0 & PETERSON & 9455 & 2003 Jul 19 \\
142860 & $\gamma$ Ser&STIS&E230H&2374-2646 & 114000 & 1822.0 & BOESGAARD & 8770 & 2001 Aug 31 \\
16895  &13 Per&   STIS & E230H & 2374-2646 & 114000 & 1979.0 & BOESGAARD & 8770 & 2000 Dec 12 \\
157214 &72 Her & STIS  & E230H & 2374-2646 & 114000 & 1870.0 & BOESGAARD & 8770 & 2001 Mar 1\\
165908 &99 Her  & STIS & E230H & 2374-2646 & 114000 & 1870.0 & BOESGAARD & 8770 & 2000 Nov 3 \\
99028  &$\iota$ Leo&STIS&E230H& 2374-2646 & 114000 & 1824.0 & BOESGAARD & 8770 & 2001 Feb 8\\
15798  &$\sigma$ Cet&STIS&E230H& 2374-2646 & 114000 & 1822.0 & BOESGAARD & 8770 & 2000 Dec 18\\
\enddata
\tablenotetext{a}{A supplemental table to Table~2 in \citet{red02}.}
\end{deluxetable}
\end{center}

\clearpage
\begin{deluxetable}{llccccl}
\tablewidth{0pt}
\tabletypesize{\tiny}
\tablecaption{Fit Parameters for \ion{Mg}{2} ISM Velocity Components\tablenotemark{a} \label{sw_table9}}
\tablehead{ HD & Other & Component & $v$ & $b$ & $\log N_{\rm MgII}$ & Reference \\
\# & Name &\# & (km s$^{-1}$) & (km s$^{-1}$) & $\log (\rm cm^{-2})$ & }
\startdata
48915\tablenotemark{b}  & $\alpha$ CMa A & 1 & 17.6   $\pm$ 1.5 & \nodata & 12.23$^{+0.02}_{-0.03}$ & 1 \\
       &                & 2 & 11.7   $\pm$ 1.5 & \nodata & 12.00$^{+0.04}_{-0.05}$ & 1 \\
%48915  & $\alpha$ CMa A & 1 & 17.6   $\pm$ 1.5 & 2.39$^{+0.10}_{-0.16}$ & 12.23$^{+0.02}_{-0.03}$ & 1 \\
%       &                & 2 & 11.7   $\pm$ 1.5 & 3.06$^{+0.35}_{-0.29}$ & 12.00$^{+0.04}_{-0.05}$ & 1 \\
103095 &CF UMa          & 1 & 2.05 $\pm$ 0.07 & 2.08 $\pm$ 0.20 & 12.58 $\pm$ 0.14 & 2 \\
157466 &SAO 85045       & 1 & --19.01 $\pm$ 0.08 & 2.40 $\pm$ 0.27 & 14.15 $\pm$ 0.59 & 2 \\
184499 &SAO 68491       & 1 & --14.00 $\pm$ 0.58 & 3.22 $\pm$ 0.38 & 13.68 $\pm$ 0.35 & 2 \\
       &                & 2 & --19.80 $\pm$ 0.82 & 3.71 $\pm$ 0.60 & 12.94 $\pm$ 0.27 & 2 \\
       &                & 3 & --27.85 $\pm$ 0.35 & 1.93 $\pm$ 0.48 & 11.97 $\pm$ 0.09 & 2 \\
30738\tablenotemark{c}  & SAO 94162      & 1 & 20.3   $\pm$ 0.4 & 3.8  $\pm$ 0.5  & 12.48 $\pm$ 0.10 & 3 \\
\enddata
\tablenotetext{a}{A supplemental table to Table~3 in \citet{red02}.}
\tablenotetext{b}{Not originally included in \citet{red02}, so listed here.}
\tablenotetext{c}{``Other Name'' listed incorrectly in \citet{red02}, so corrected here.}
\tablerefs{(1) H\'{e}brard et al. 1999; (2) this paper; (3) Redfield \& Linsky 2001.}
\end{deluxetable}  

\clearpage
\begin{deluxetable}{llccccl}
\tablewidth{0pt}
\tabletypesize{\tiny}
\tablecaption{Fit Parameters for \ion{Fe}{2} ISM Velocity Components\tablenotemark{a} \label{sw_table10}}
\tablehead{ HD & Other & Component & $v$ & $b$ & $\log N_{\rm FeII}$ & Reference \\
\# & Name &\# & (km s$^{-1}$) & (km s$^{-1}$) & $\log (\rm cm^{-2})$ & }
\startdata
48915\tablenotemark{b}  & $\alpha$ CMa A & 1 & 17.6   $\pm$ 1.5 & \nodata & 11.94$^{+0.01}_{-0.02}$ & 1 \\
       &                & 2 & 11.7   $\pm$ 1.5 & \nodata & 11.74 $\pm$ 0.03 & 1 \\
%48915  & $\alpha$ CMa A & 1 & 17.6   $\pm$ 1.5 & 1.62$^{+0.10}_{-0.13}$ & 11.94$^{+0.01}_{-0.02}$ & 1 \\
%       &                & 2 & 11.7   $\pm$ 1.5 & 2.86$^{+0.30}_{-0.29}$ & 11.74 $\pm$ 0.03 & 1 \\
61421\tablenotemark{c}  & $\alpha$ CMi        & 1 & 22.0  $\pm$ 0.10& 2.1 $\pm$ 0.10 & 11.94 $\pm$ 0.03 & 2 \\
       &                & 2 & 19.0  $\pm$ 0.10& 2.1 $\pm$ 0.10 & 12.05 $\pm$ 0.02 & 2 \\
4614 & $\eta$ Cas A     & 1 & 11.18 $\pm$ 0.40 & 3.49 $\pm$ 0.43 & 12.43 $\pm$ 0.09 & 3 \\
     &                  & 2 &  5.80 $\pm$ 0.87 & 2.00 $\pm$ 0.89 & 11.67 $\pm$ 0.24 & 3 \\
216956\tablenotemark{d} & $\alpha$ PsA   & 1 & --3.2  $\pm$ 1.5 & 2.7             & 12.58            & 4 \\
       &                & 2 & --9.7  $\pm$ 1.5 & 2.8             & 12.26            & 4 \\
\enddata
\tablenotetext{a}{A supplemental table to Table~4 in \citet{red02}.}
\tablenotetext{b}{Not originally included in \citet{red02}, so listed here.}
\tablenotetext{c}{One-component fit mistakenly listed in \citet{red02}, preferred two-component fit given here.}
\tablenotetext{d}{Missing line of table in \citet{red02} confused fits for these two sightlines, corrected here.}
\tablerefs{(1) H\'{e}brard et al. 1999; (2) Linsky et al 1995; (3) this paper; (4) Ferlet et al 1995; (5) Lallement et al 1995.}
%(3) Ferlet et al 1995; (4) Lallement et al 1995; (5) this paper.}
\end{deluxetable}

\end{document}